\documentclass[aps,prd,notitlepage,reprint,twocolumn,showpacs,superscriptaddress,longbibliography,nofootinbib,floatfix]{revtex4-1}

\usepackage{amsmath,amssymb,amsfonts}
\usepackage{graphicx}
\usepackage{verbatim}
\usepackage{pifont}

\usepackage{hyperref}
\usepackage{slashed}
\usepackage{verbatim}

\usepackage[usenames]{color}
\definecolor{purple}{rgb}{0.5,0,0.8}

\newcommand{\be}{\begin{equation}}
\newcommand{\ee}{\end{equation}}
\newcommand{\bi}{\begin{itemize}}
\newcommand{\ei}{\end{itemize}}
\newcommand{\bea}{\begin{eqnarray}}
\newcommand{\eea}{\end{eqnarray}}

\newcommand{\ud}{\mathrm{d}}

\newcommand{\LCm}{{\scriptscriptstyle -}} 
\newcommand{\LCp}{{\scriptscriptstyle +}}

\newcommand{\LCperp}{{\scriptscriptstyle \perp}}

\newcommand{\LCpara}{{\scriptscriptstyle \parallel}}

\usepackage[T1]{fontenc} 

\usepackage{bm}

\begin{document}

\title{Perturbative methods for assisted nonperturbative pair production}

\author{Greger Torgrimsson}
\email{greger.torgrimsson@uni-jena.de}
\affiliation{Theoretisch-Physikalisches Institut, Abbe Center of Photonics,
	Friedrich-Schiller-Universit\"at Jena, Max-Wien-Platz 1, D-07743 Jena, Germany}
\affiliation{Helmholtz Institute Jena, Fr\"obelstieg 3, D-07743 Jena, Germany}

\begin{abstract}
In the dynamically assisted Schwinger mechanism, the pair production probability is significantly enhanced by including a weak, rapidly varying field in addition to a strong, slowly varying field. 
In a previous paper we showed that several features of dynamical assistance can be understood by a perturbative treatment of the weak field. Here we show how to calculate the prefactors of the higher-order terms, which is important because the dominant contribution can come from higher orders. We give a new and independent derivation of the momentum spectrum using the worldline formalism, and extend our WKB approach to calculate the amplitude to higher orders. We show that these methods are also applicable to doubly assisted pair production.
\end{abstract}
\maketitle

\section{Introduction}

Schwinger pair production~\cite{Sauter:1931zz,Heisenberg:1935qt,Schwinger:1951nm}	
by a slowly varying electric field will probably not be observed in the near future, as the probability is too small even for the highest intensities that will be available. However, by adding to the slowly varying field a weaker, but rapidly varying field, one can increase the probability by orders of magnitude~\cite{Schutzhold:2008pz,Orthaber:2011cm,Otto:2014ssa,Linder:2015vta,Torgrimsson:2017pzs,Torgrimsson:2017cyb,Aleksandrov:2018uqb}, and hence significantly reduce the required field strength. One key aspect of Schwinger pair production is its nonperturbative dependence on the field strength.  
When adding assisting, high-frequency fields, one might like to have a probability that is still nonperturbative in the field strength, as such high-frequency fields can lead to perturbative pair production, which could be produced in experiments similar to the famous one at SLAC~\cite{Bamber:1999zt}. This does not mean, though, that the probability has to be nonperturbative in both fields separately. Indeed, in our previous paper~\cite{Torgrimsson:2017pzs} we showed that the weak field can in many cases be treated perturbatively, which allows us to find explicit analytical expressions to study dynamical assistance for a large class of fields. 

Let us first recall some of the most important results in~\cite{Torgrimsson:2017pzs}. Consider a time-dependent electric field given by $E_z(t)=E(f_0(t)+\varepsilon f(t))$, where $E\ll1$ is the field strength of the strong field and $f$ the field shape of the weak field, with $\varepsilon\ll1$. We assume that the weak field is much faster than the strong field and in most of the calculations we can set $f_0\approx1$. 
We use units with $\hbar=c=1$ as well as $m=1$, where $m$ is the electron mass, and absorb a factor of the charge into the definition of the background field $eE\to E$. For example, Schwinger's critical field is in these conventions simply $E_{\rm crit}=m^2/e=1$.  
In~\cite{Torgrimsson:2017pzs} we expanded the pair production probability as 
\be
P_{e^\LCp e^\LCm}=P_0+\varepsilon P_1+\varepsilon^2 P_2+\dots \;, 
\ee
where $P_0\sim\exp(-\pi/E)$ gives the ordinary Schwinger pair production probability~\cite{Sauter:1931zz,Heisenberg:1935qt,Schwinger:1951nm}, and the higher-order terms give dynamical assistance. Despite being suppressed by higher powers of $\varepsilon$, in regimes with significant dynamical assistance the contribution from these higher orders is much larger than $P_0$ thanks to the exponential enhancement due to photon absorption. 
  
By expressing the weak field in terms of its Fourier transform we found $P_N$ in terms of $N$ Fourier integrals,
\be\label{PNFourierExpansion} 
P_N=\int\ud\omega_1\dots\ud\omega_N f(\omega_1)\dots f(\omega_N) F_N \;,
\ee  
where $\omega_i$ are the Fourier frequencies. $F_N$ contains $\delta(\omega_1+\dots+\omega_N)$ for a constant strong field. The dominant contribution to the integrand is given by~\cite{Torgrimsson:2017pzs}
\be\label{PNintermsofSigma} 
F_N\sim\exp\left\{-\frac{2}{E}\left(\arccos\Sigma-\Sigma\sqrt{1-\Sigma^2}\right)\right\} \;,
\ee 
where
\be
\Sigma=\frac{1}{2}\sum_{i=1}^{J}\omega_i 
\ee
is the sum of the positive frequencies, ordered for simplicity such that $\omega_i>0$ for $1\leq i\leq J<N$ for some $J$, divided by the energy of a real pair at rest. For even $N$, the dominant contribution comes from $J=N/2$, where half of the $\omega_i$'s are positive and the other half negative. 

For a Sauter pulse, $\propto\text{sech}^2(\omega t)$, the Fourier transform scales as $f(\omega_1)\sim\exp(-|\omega_1|/\omega_*)$ for 
$|\omega_1|\gg\omega_*$, where $\omega_*=2\omega/\pi$. We focus on $|\omega_1|\gg\omega_*$ because that is the part of the Fourier integrals which gives the dominant contribution.    
By performing the Fourier integrals in~\eqref{PNFourierExpansion} we found~\cite{Torgrimsson:2017pzs}    
\be\label{SauterUniversalExponent}
P\sim\exp\left\{-\frac{2}{E}\left(\frac{\sqrt{\gamma_*^2-1}}{\gamma_*^2}+\arcsin\frac{1}{\gamma_*}\right)\right\} \;,    
\ee
where the normalized Keldysh parameter is given by $\gamma_*=\gamma/\gamma_{\rm crit}$, $\gamma=\omega/E$ and for a Sauter pulse $\gamma_{\rm crit}=\pi/2$. For a Sauter pulse, \eqref{SauterUniversalExponent} gives the exponential scaling of $P_N$ for all $N>1$, 
which, since the higher orders are suppressed by $\varepsilon^N$, means that already $\varepsilon^2 P_2$ gives the dominant contribution (independently of $\gamma$ and $E$),
and the exponent agrees exactly with the exponent found in~\cite{Schutzhold:2008pz} by treating both the strong, constant field and the weak, Sauter pulse with nonperturbative methods. On a conceptual level, this tells us that the dependence on the weak field is perturbative, which might not be obvious in other approaches. On a practical level, the fact that the dominant contribution is already given by $\varepsilon^2 P_2$ allows us to find analytical expressions for the prefactor too, which we have shown agree well with the exact numerical result~\cite{Torgrimsson:2017pzs}. This has the advantage of working also for other fields with similar Fourier transforms at large frequencies.

In contrast, for a Gaussian pulse and for a monochromatic field, we found that $P_N$ increases as one goes to higher orders. Because of the factor of $\varepsilon^N$ in the prefactor, there is in general a dominant order~\cite{Torgrimsson:2017pzs}, i.e. the order $N_{\rm dom}$ which gives the dominant contribution, which in this case can be $N_{\rm dom}>2$. By treating $N$ as a continuous variable we found~\cite{Torgrimsson:2017pzs}
\be\label{domNgauss}
N_{\rm dom}^{\rm Gauss}\sim\frac{2}{E|\ln\varepsilon|}\frac{\sqrt{\chi^2-1}}{\chi^2} \;,
\ee
where $\chi\sim\gamma/\sqrt{|\ln\varepsilon|}\sim\gamma/\gamma_{\rm crit}$,
and by estimating the sum of all orders with this ``saddle point'' for $N$, we recover~\eqref{SauterUniversalExponent}, but with $\gamma_{\rm crit}\sim\sqrt{|\ln\varepsilon|}$ for a Gaussian pulse (and $\gamma_{\rm crit}\sim|\ln\varepsilon|$ for a monochromatic field), which agrees with the $\gamma_{\rm crit}$ found previously in~\cite{Linder:2015vta}.
From~\eqref{domNgauss} we see that below the threshold ($\gamma<\gamma_{\rm crit}\sim\sqrt{|\ln\varepsilon|}$) the dominant order is zero, which is natural since there is no exponential enhancement of the higher orders there and so one basically has an ordinary power series. As $\gamma$ increases the dominant order first increases, and then it reaches a maximum after which it decreases, which is also natural since at sufficiently high frequencies already the first order can provide enough energy to give the dominant contribution. The maximum dominant order is at $\chi=\sqrt{2}$, which is also the most interesting region, because there one can expect the maximum enhancement compared to both pure Schwinger and purely perturbative pair production. Apart from the weak, logarithmic dependence on $\varepsilon$, we see that the most important parameter determining the dominant order is the field strength $E$. Weaker $E$ leads to a higher dominant order, which is illustrated in Fig.~3 in~\cite{Torgrimsson:2017pzs}. 

Let us put these results into a bigger picture.
Consider pair production in an ensemble of constant energy $\mathcal{E}$. The exponential part of the probability for this process was derived in~\cite{Gould:2017fve} (see Eq.~(66) in~\cite{Gould:2017fve}), which to leading order\footnote{The results in~\cite{Gould:2017fve} also contains higher orders in $\alpha$, which can be seen as an invitation to consider such higher orders also in our case.} in $\alpha$ can be expressed as
\be\label{PmathcalE}
P\sim\exp\bigg\{-\frac{2}{E}\bigg(\text{arccos}\frac{\mathcal{E}}{2}-\frac{\mathcal{E}}{2}\sqrt{1-\left(\frac{\mathcal{E}}{2}\right)^2}\bigg)\bigg\} \;.
\ee
By identifying our sum over ``absorbed'' Fourier frequencies $\sum\omega_i$ in~\eqref{PNintermsofSigma} with the energy $\mathcal{E}$ in~\eqref{PmathcalE} we find an exact agreement. 
As an aside, we note that the constant energy result in~\cite{Gould:2017fve} was obtained by a Legendre transform of a corresponding result for constant temperature $T$, which has exactly the same functional form as the exponential in~\eqref{SauterUniversalExponent} for a Sauter pulse, but with $\gamma_*\to2mT/(qE)$, see also~\cite{Brown:2015kgj}. We can understand this as being due to the fact that the exponential scaling of the Fourier transform of a Sauter pulse effectively acts as a Boltzmann factor, and so performing the Fourier integrals with the saddle-point method effectively corresponds to doing the Legendre transform in~\cite{Gould:2017fve} in reverse.  

Many aspects in Schwinger pair production have close analogies in tunneling in semiconductors~\cite{Linder:2015fba}. In particular, dynamically assisted Schwinger pair production is analogous to the Franz-Keldysh effect~\cite{Linder:2015fba,FranzFranzKeldysh,KeldyshFranzKeldysh,FranzKeldyshBook}, where semiconductor tunneling in an electric field is assisted by higher-frequency photons.
The Franz-Keldysh effect in QED was very recently studied in~\cite{Taya:2018eng}. There exists certain replacement rules~\cite{Linder:2015fba} for translating results for semiconductor tunneling to Schwinger pair production or vice versa. To translate our result~\eqref{PNintermsofSigma} for Schwinger pair production to the corresponding result for semiconductor tunneling we have to replace~\cite{Linder:2015fba} $qE_{\rm crit}\to c_*^3m_*^2$ and $\Sigma\to\omega/(2m_*c_*^2)$, where $c_*$ and $m_*$ are semiconductor parameters related to the effective speed of light and the band gap. The resulting exponential agrees exactly with the literature on the Franz-Keldysh effect, see Eq.~(32) in~\cite{AronovPikusJETP24-339} or Eq.~(C11) in~\cite{WeilerPhotonAssistedTunneling} for the first order, and~\cite{GarciaNonlinearFranzKeldysh} for higher orders.
We will study this analogy further elsewhere~\cite{FranzKeldysToAppear}.   
 
Of course, this does not mean that we can obtain all our results by just replacing various parameters in existing literature results. 
In particular, this does not tell us how different field shapes affect the probability or how to obtain the prefactor.  

This paper is organized as follows.
In~\cite{Torgrimsson:2017pzs} we calculated the prefactor of the momentum spectrum using a WKB approach; here in Sec.~\ref{Momentum spectrum from the worldline formalism} we rederive those results using a completely different approach, namely one based on the worldline formalism. In~\cite{Torgrimsson:2017pzs} we calculated the exponential part of the probability to all orders, but the prefactor only up to $N=2$; here in Sec.~\ref{Using the propagator in a constant electric field} we show how to calculate the prefactor at higher orders and give examples where we go up to $N=6$. In~\cite{Torgrimsson:2017pzs} we showed that $N=2$ is in general enough for Sauter-like fields but not always enough for a Gaussian field, and we gave an example where $N=2$ is not enough for a Gaussian field; here in Sec.~\ref{Using the propagator in a constant electric field} we show that going to $N=4$ does give a good agreement for that example, which is hence an explicit example, with the prefactor included, where the dominant order is higher than two. In~\cite{Torgrimsson:2017pzs} we calculated the exponentials at higher order using the worldline formalism; here in Sec.~\ref{Higher orders using the propagator} we show how to obtain these using the WKB approach.
In Sec.~\ref{HigherOrderPreWorldline} we show how the results in Sec.~\ref{Using the propagator in a constant electric field} for the higher-order prefactors of the integrated probability can be obtained by including the prefactor in the worldline approach we used in~\cite{Torgrimsson:2017pzs}. 
In~\cite{Torgrimsson:2016ant} we introduced a doubly assisted mechanism, where Schwinger pair production is assisted by both a weak (coherent) field and a single, on-shell high-energy photon, which we studied by treating both the strong and the weak field with nonperturbative methods; here in Sec.~\ref{Double assistance section} we study this mechanism by treating the weak field perturbatively, which offers the possibility to obtain the prefactor e.g. for Sauter-like weak fields.

\section{Momentum spectrum from the worldline formalism}\label{Momentum spectrum from the worldline formalism}

\begin{figure}
\includegraphics[width=\linewidth]{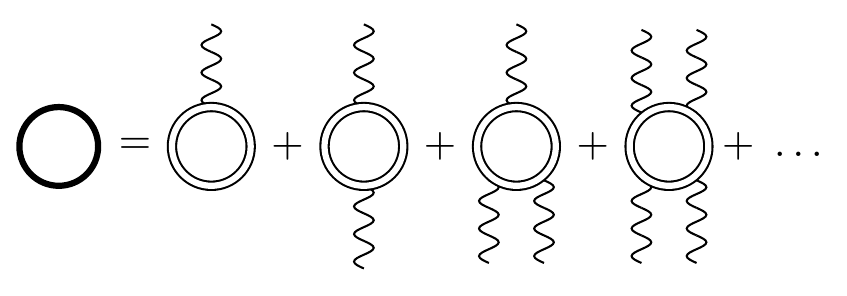}
\caption{The expansion of the one-loop effective action in terms of the weak field. The bold line represents fermions dressed by both the strong and the weak field; the double lines represent fermions dressed by the strong field alone; and the wiggly lines represent photons from the weak field (these photons are off-shell for the fields we focus on here).}
\label{GammaExpansion1-fig}
\end{figure}

In this section we rederive the momentum spectrum of the produced particles using the worldline-momentum representation of the effective action~\cite{Dumlu:2011cc}. To the best of our knowledge this formalism\footnote{Note, though, that a similar representation of the propagator was used in~\cite{Barut:1989mc}.} has so far only been used in~\cite{Dumlu:2011cc}, but we show here that it offers a useful alternative to the WKB approach for obtaining the momentum spectrum, including the prefactor.  
The pair production probability is given by the imaginary part of the effective action $P_{e^+e^-}=2\text{Im }\Gamma$, which in turn is given in the usual worldline representation by (see e.g.~\cite{Affleck:1981bma,Dunne:2005sx,Dunne:2006st})
\be\label{GammaUsual}
\Gamma=2\int_0^\infty\frac{\ud T}{T}\oint\mathcal{D}x\text{ spin }e^{-i\left(\frac{T}{2}+\int_0^1\frac{\dot{x}^2}{2T}+A\dot{x}\right)} \;,
\ee  
where $x^\mu(0)=x^\mu(1)$ and the spin factor is in general given by the trace of a path-ordered exponential
\be
\text{spin}=\frac{1}{4}\text{tr ``path order''}\exp\left\{-\frac{i T}{4}\int_0^1\sigma^{\mu\nu}F_{\mu\nu}\right\} \;,
\ee
but, for the one-component fields we consider here, $A_\mu=\delta_\mu^3A_3(t)$, it reduces to~\cite{Dunne:2005sx,Dumlu:2011cc}
\be\label{simpleSpin}
\text{spin}=\cos\left(\frac{iT}{2}\int_0^1A_3'(t)\right) \;.
\ee
The standard representation~\eqref{GammaUsual} gives the total/integrated probability. To obtain the spectrum, we follow~\cite{Dumlu:2011cc} and rewrite the effective action in a momentum representation as
\be\label{worldline-spectrum-start}
\begin{split}
\Gamma=2V_3&\int\frac{\ud^3p}{(2\pi)^3}\int_0^\infty\frac{\ud T}{T}\oint\mathcal{D}t\text{ spin } \\
&\exp\left\{-i\left(\frac{Tm_\LCperp^2}{2}+\int_0^1\frac{\dot{t}^2}{2T}+\frac{T}{2}(p_3-A)^2\right)\right\} \;,
\end{split}
\ee
where $m_\LCperp=\sqrt{1+p_\LCperp^2}$, $p_\LCperp=\{p_1,p_2\}$, and
where the integrand of the ${\bf p}$-integral gives the momentum spectrum\footnote{The effective action gives, of course, the probability of producing any number of pairs, but this is approximately equal to the probability of producing a single pair.}.
We consider a strong constant field $E$ plus a weak, rapidly varying field $a(t)$, $A_3=Et+a(t)$, and expand~\eqref{worldline-spectrum-start} in the weak field $a\sim\varepsilon\ll1$
\be\label{Gamma012dots}
\Gamma=\Gamma_0+\varepsilon\Gamma_1+\varepsilon^2\Gamma_2+\dots
\ee
This expansion is illustrated in Fig.~\ref{GammaExpansion1-fig}.
After expressing the weak field in terms of its Fourier transform,
\be\label{aFourierDefinition}
a(t)=\int\frac{\ud\omega_1}{2\pi} e^{-i\omega_1t}a(\omega_1) \;,
\ee 
we find Gaussian path integrals which we can perform with methods similar to those used in~\cite{Shaisultanov:1995tm,Schubert:2000yt,Schubert:2001he} to calculate N-photon amplitudes in constant background fields.

Two typical fields are the Sauter and the Gaussian pulse. The Sauter pulse is given by
\be\label{SauterDef}
\begin{split}
	&a(t)=\frac{E\varepsilon}{\omega}\tanh\omega t \qquad\to \\
	&a(\omega_1)=\frac{E\varepsilon}{\omega^2}\frac{\pi i}{\sinh\frac{\pi\omega_1}{2\omega}}\approx\text{sign}(\omega_1)2\pi i\frac{E\varepsilon}{\omega^2}e^{-\frac{|\omega_1|}{\omega_*}} \;,
\end{split}
\ee
where we have introduced $\omega_*=2\omega/\pi$ to make it easier to generalize to other fields that have exponentially decaying Fourier transforms for Fourier frequencies above the characteristic frequency, i.e. $|\omega_1|\gg\omega$ (recall that this gives the dominant contribution). The Gaussian pulse is given by
\be\label{WeakGaussianDef}
\begin{split}
	&a(t)=\frac{E\varepsilon}{\omega}\frac{\sqrt{\pi}}{2}\text{erf}(\omega t) 
	\qquad\to \\
	&a(\omega_1)=\frac{E\varepsilon}{\omega}\frac{i\sqrt{\pi}}{\omega_1}e^{-\left[\frac{\omega_1}{2\omega}\right]^2} \;.
\end{split}
\ee

\subsection{Zeroth order $\Gamma_0$}

We begin with $\Gamma_0$. This gives of course the well-known constant field result~\cite{Sauter:1931zz,Heisenberg:1935qt,Schwinger:1951nm}, but it allows us to check the overall normalization constant, which is the same for the higher orders. Changing from Minkowski to Euclidean variables
\be\label{Tt-change}
T\to-iT \qquad t\to-it+\frac{p_3}{E}
\ee
gives us
\be\label{Gamma0Euclidean}
\begin{split}
\Gamma_0=2iV_3&\int\frac{\ud^3p}{(2\pi)^3}\int_0^\infty\frac{\ud T}{T}\oint\mathcal{D}t\cos\frac{ET}{2} \\
&\exp-\left(\frac{Tm_\LCperp^2}{2}+\int_0^1\frac{\dot{t}^2}{2T}-\frac{T}{2}(Et)^2\right) \;.
\end{split}
\ee
We separate the center of mass $t_0$ from the time variable $t(\tau)\to t_0+t(\tau)$, Fourier expand
\be\label{tFourierExpand}
t(\tau)=\sum\limits_{n=1}^\infty a_n\cos2\pi n\tau+b_n\sin2\pi n\tau 
\ee 
and calculate the path integral by multiplying together all the eigenvalues. The path integral is normalized according to
\be
\oint\mathcal{D}t\exp-\int_0^1\frac{\dot{t}^2}{2T}=\frac{1}{\sqrt{2\pi T}} \;,
\ee
so, by dividing by the free integral (cf.~\cite{Affleck:1981bma}), we obtain 
\be\label{tpathbeta}
\begin{split}
&\int\ud t_0\oint\mathcal{D}t\exp\left\{-\frac{1}{2T}\int t[-\partial^2-(ET)^2]t\right\} \\
&=\frac{1}{\sqrt{2\pi T}}i\frac{\sqrt{2\pi T}}{ET}\prod\limits_{n=1}^\infty\frac{(2\pi n)^2}{(2\pi n)^2-(ET)^2}=\frac{i}{2\sin s} \;,
\end{split}
\ee
where $s=ET/2$ and the product can be obtained e.g. from Eq.~(1.431.1) in~\cite{GradshteynRyzhik}. 
The integration contour for $s$ goes over the poles and gives an imaginary part to the effective action. To leading order we find
\be\label{ImGamma0fin}
\begin{split}
2\text{Im }\Gamma_0&=-2\text{Im }V_3\int\frac{\ud^3p}{(2\pi)^3}\int_0^\infty\frac{\ud s}{s}\cot s\; e^{-sm_\LCperp^2/E} \\
&\approx2V_3\int\frac{\ud^3p}{(2\pi)^3}e^{-\pi m_\LCperp^2/E}= V_4\frac{E^2}{4\pi^3}e^{-\frac{\pi}{E}} \;.
\end{split}
\ee 
This is of course the leading term in the well-known Schwinger formula. We can thus confirm that the normalization factor in~\eqref{Gamma0Euclidean} is correct.

\subsection{First order $\Gamma_1$}\label{First order Gamma1}

The first order $\Gamma_1$ corresponds to the cross term between the zeroth and first order amplitudes, $2\text{Re }\frak{A}_0^*\frak{A}_1$, which we calculated in~\cite{Torgrimsson:2017pzs} using a WKB approach.
Here we find by expanding~\eqref{worldline-spectrum-start}
\be\begin{split}
\varepsilon\Gamma_1=&2V_3\int\frac{\ud^3p}{(2\pi)^3}\int_0^\infty\frac{\ud T}{T}\oint\mathcal{D}t\cos\frac{iET}{2} \\
&\exp\left\{-i\left(\frac{Tm_\LCperp^2}{2}+\int_0^1\frac{\dot{t}^2}{2T}+\frac{T}{2}(p_3-Et)^2\right)\right\} \\
&\int\frac{\ud\omega_1}{2\pi}a(\omega_1)\int_0^1\ud\tau_1\frac{iT}{2}e^{-i\omega_1t(\tau_1)} \\
&\left(i\omega_1\tan\frac{iET}{2}+2[p_3-Et(\tau_1)]\right) \;,
\end{split}
\ee
where the first two lines are the same as for $\Gamma_0$ and hence have the same normalization.
We change to Euclidean variables according to~\eqref{Tt-change}.

To make the exponent quadratic in $t$ we make a replacement $t\to t_\text{cl}+t$. Since the ``classical'' solution $t_\text{cl}$ takes the same form for all orders, $\Gamma_N$, we consider temporarily general $N$. We find $t_\text{cl}$ by expanding its equation of motion, 
\be\label{tcleom1}
(\partial^2+(ET)^2)t_\text{cl}(\tau)=T\sum\limits_{i=1}^N\omega_i\delta(\tau-\tau_i) \;,
\ee 
in terms of Fourier modes, which yields
\be\label{tclsum1}
\begin{split}
t_\text{cl}(\tau)&=T\sum\limits_{i=1}^N\omega_i\sum\limits_{n=-\infty}^\infty\frac{e^{2\pi i n(\tau-\tau_i)}}{(ET)^2-(2\pi n)^2} \\
&=\frac{1}{2E}\sum_i\omega_i\frac{\cos[s(1-2|\tau-\tau_i|)]}{\sin s} \;,
\end{split}
\ee
where the sum over $n$ can be performed using Eq.~(1.445.2) or~(1.445.9) in~\cite{GradshteynRyzhik}. 
With the linear term removed from the exponent, the $t$-integral is now the same for all orders and is given by~\eqref{tpathbeta}.

Returning to $N=1$, the $\tau_1$-integral is trivial and we find
\be\label{sIntGamma1}
\begin{split}
\varepsilon\Gamma_1=-V_3&\int\frac{\ud^3p}{(2\pi)^3}\int_0^\infty\frac{\ud s}{s}\cot s\; e^{-\frac{s m_\LCperp^2}{E}} \\
i&\int\frac{\ud\omega_1}{2\pi}a(\omega_1)\frac{s}{E}\frac{\omega_1}{\sin s\cos s}e^{-\frac{1}{E}\big(ip_3\omega_1+\frac{\omega_1^2}{4}\cot s\big)} \;.
\end{split}
\ee
Performing this $p_3$ integral simply gives a delta function $\delta(\omega_1)$ which reduces the exponential in $\Gamma_1$ to the constant field case, and then there is nothing to compensate for the small prefactor, $a\ll1$, which means that $\Gamma_1$ only gives a small correction to the integrated probability. 
Note though that this delta function does not automatically make the prefactor zero, since
$-i\omega_1a(\omega_1)|_{\omega_1=0}=\int\ud t a'(t)$,
which can be nonzero depending on how the total field is separated into a strong and a weak field\footnote{To recover the prefactor obtained by replacing $E\to E+\int a'/V_0$ in the constant field result, the last expression in~\eqref{ImGamma0fin}, and expanding in $a'$, one has to remember that converting the $p_3$-integral into a volume factor also leads to a field-dependent factor.}. In any case, we are not interested here in such small corrections to the constant field result. We are instead interested in higher-order terms that come from nonzero Fourier frequencies and that, due to exponential enhancement, can be much larger than the zeroth order/constant field probability.
While $\varepsilon\Gamma_1$ gives a negligible contribution to the integrated probability, it can give important interference effects in the spectrum. 
   
We perform the proper-time $s$ integral with the saddle-point method. We define for convenience $\Sigma=|\omega_1|/(2m_\LCperp)$. The saddle-point equation $\sin^2s=\Sigma^2$ has two solutions in the region $0<s<\pi$. Although the first saddle point $s=\arcsin\Sigma$ ($0<s<\pi/2$) gives a larger exponential, the Gaussian integral around it is real so, since the Fourier integral is also real, this saddle point does not contribute to the imaginary part of the effective action. Thus, only the second saddle point 
\be\label{sSaddleGamma1}
s_*=\frac{\pi}{2}+\arccos\Sigma
\ee
($\pi/2<s_*<\pi$) is relevant here. Let $\delta s=s-s_*$ be the perturbation around this saddle point, then for small $\delta s$ the exponent is given by 
$\exp\left\{\frac{m_\LCperp^2}{E}\frac{\sqrt{1-\Sigma^2}}{\Sigma}\delta s^2\right\}$.
The first part of the integration contour follows the real axis from $s=0$ to the saddle point~\eqref{sSaddleGamma1} and gives a purely real contribution to the integral. The second part of the contour starts at the saddle point and follows the steepest descent where the imaginary part of the exponent is zero. Since the second part starts perpendicular to the real axis, it gives us an imaginary contribution to $\Gamma$. Recalling that the initial contour followed what now corresponds to the imaginary axis, we have $\ud s\propto+i$ near the saddle point. The Gaussian integral around this saddle point hence gives
%\be\label{sSaddleIntegralGamma1}
%\begin{split}
%\int\ud\delta s\to&\frac{i}{2}\left[\frac{\pi E}{m_\LCperp^2}\frac{\Sigma}{\sqrt{1-\Sigma^2}}\right]^\frac{1}{2}e^{-\frac{m_\LCperp^2}{E}\left(\frac{\pi}{2}+\arccos\Sigma-\Sigma\sqrt{1-\Sigma^2}\right)} \\
%&+\text{``something real''} \;,
%\end{split}
%\ee
\be\label{sSaddleIntegralGamma1}
\int\ud s f(s)=\frac{i}{2}\left[\frac{\pi E}{m_\LCperp^2}\frac{\Sigma}{\sqrt{1-\Sigma^2}}\right]^\frac{1}{2}\!f(s_*)
+\text{``something real''} \;,
\ee
where a factor of $1/2$ comes from having only half of a Gaussian integral. 
Collecting all the terms we find
\be\label{finalGeneralGamma1}
\begin{split}
2\text{Im }&\varepsilon\Gamma_1=2V_3\int\frac{\ud^3p}{(2\pi)^3}2\text{Re}\int_0^\infty\frac{\ud\omega_1}{2\pi}\sqrt{\frac{\pi}{E}}\frac{(-i)a(\omega_1)}{\sqrt{\Sigma}(1-\Sigma^2)^\frac{1}{4}} \\
&\exp\left\{\!-\frac{m_\LCperp^2}{E}\left[\frac{\pi}{2}+2iP\Sigma+\arccos\Sigma-\Sigma\sqrt{1-\Sigma^2}\right]\right\} \;,
\end{split}
\ee
where $P=p_3/m_\LCperp$. Clearly, the saddle-point method that we have used to derive~\eqref{finalGeneralGamma1} is only valid for $0<\Sigma<1$ or $0<|\omega_1|<2m_\LCperp$. Fortunately, the $\omega_1$~integral has in general a saddle point in this range, and we are interested in regimes where the dominant contribution comes from such saddle points. So, the integration limits should in fact be restricted to regions that are sufficiently close to the saddle points, but we do not explicitly write out these integration limits. The same holds for other Fourier integrals below.

\begin{figure}
\includegraphics[width=\linewidth]{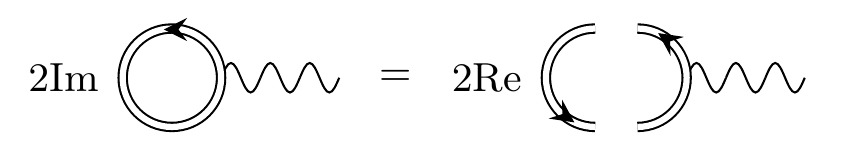}
\caption{A diagrammatic illustration of~\eqref{Gamma1Amp0Amp1momentum}. One of the diagrams on the right-hand side represents the complex conjugate of the corresponding amplitude.}
\label{Gamma1Amp0Amp1-fig}
\end{figure}

To compare~\eqref{finalGeneralGamma1} with our results in~\cite{Torgrimsson:2017pzs}, we first recall that in~\cite{Torgrimsson:2017pzs} the momentum spectrum was obtained from the amplitude, $\frak{A}$, as  
\be
P_{e^\LCp e^\LCm}=V_3\int\frac{\ud^3p}{(2\pi)^3}\left|\frak{A}_0+\varepsilon\frak{A}_1+\varepsilon^2\frak{A}_2+\dots\right|^2 \;,
\ee
where the zeroth order amplitude is given by
\be\label{zerothOrderA}
\frak{A}_0=\delta_{s,s'}\exp\left\{-\frac{m_\LCperp^2}{E}\frac{\pi}{2}+\frac{i m_\LCperp^2}{E}\phi(P)\right\} \;,
\ee
and, from Eqs.~(2.7), (4.14) and~(4.23) in~\cite{Torgrimsson:2017pzs}, the first order amplitude can be expressed as
\be\label{firstOrderA}
\begin{split}
\varepsilon\frak{A}_1=&\delta_{s,s'}\int_0^\infty\frac{\ud\omega_1}{2\pi}a(\omega_1)(-i)\sqrt{\frac{\pi}{E}}\frac{\exp\left\{\frac{i m_\LCperp^2}{E}\phi(P)\right\}}{\sqrt{\Sigma}(1-\Sigma^2)^\frac{1}{4}} \\
&\exp\left\{-\frac{m_\LCperp^2}{E}\left(2iP\Sigma+\arccos\Sigma-\Sigma\sqrt{1-\Sigma^2}\right)\right\} \;,
\end{split}
\ee 
where the restriction to $\omega_1>0$ is due to the fact that this gives the dominant contribution, and $\Sigma=\omega_1/(2m_\LCperp)$. Here $s$ and $s'$ describe the spin of the electron and positron, and the $\delta_{s,s'}$ means that the sum over spins simply gives a factor of $2$ (the phase $i\phi(P)$ is completely irrelevant and is due to an arbitrary choice in the WKB solutions). Thus, we find perfect agreement between the worldline-momentum and the WKB approach, i.e.
\be\label{Gamma1Amp0Amp1momentum}
2\text{Im }\varepsilon\Gamma_1=V_3\int\frac{\ud^3p}{(2\pi)^3}\sum_\text{spin}2\text{Re }\frak{A}_0^*\varepsilon\frak{A}_1 \;,
\ee
where $\Gamma_1$, $\frak{A}_0$ and $\frak{A}_1$ are given by~\eqref{finalGeneralGamma1}, \eqref{zerothOrderA} and~\eqref{firstOrderA}, respectively.   
This relation is illustrated in Fig.~\ref{Gamma1Amp0Amp1-fig}.

We have demonstrated this equivalence without having to specify the shape of the weak field. To make this agreement more explicit, we consider in the next two subsections the Sauter and the Gaussian pulse.

\subsubsection{Sauter pulse} 
    
To obtain the spectrum we now only have the Fourier integral left, and to perform it we need to specify the shape of the weak field. We begin with the Sauter pulse~\eqref{SauterDef}. 
We perform the Fourier integral with the saddle-point method. There are two saddle points with opposite signs that give complex conjugate contributions. We can therefore without loss of generality focus on $\text{Re }\omega_1>0$. 
The saddle point for $\omega_1$ is given by $\Sigma(\omega_1)=\sqrt{1+\hat{\pi}_3^2}=:\hat{\pi}_0=\pi_0/m_\LCperp$, where $\hat{\pi}_3=(p_3-i/\gamma_*)/m_\LCperp$ can be thought of as the ``physical'' momentum of an electron in a constant electric field at an imaginary time, and $\gamma_*=\omega_*/E$ is the combined Keldysh parameter suitably normalized.  
Notice that this saddle point corresponds to a Fourier frequency of $\omega_1=2\pi_0$, which is on the order of the electron mass even for a characteristic frequency $\omega\ll1$. The exponential suppression of the Fourier transform at such high frequencies (we assume $\omega\ll1$) contributes to the overall exponential behavior of the pair production probability. 
Collecting everything we finally find
\be\label{Gamma1SauterFin}
2\text{Im }\varepsilon\Gamma_1=2V_3\int\frac{\ud^3p}{(2\pi)^3}2\text{Re }\frac{2\pi E\varepsilon}{\omega^2}\frac{1}{\hat{\pi}_0}e^{\!-\frac{m_\LCperp^2}{E}\left[\frac{\pi}{2}+i\phi(\hat{\pi}_3)\right]} \;,
\ee
which agrees with what we found in our previous paper~\cite{Torgrimsson:2017pzs} for the cross term between the zeroth and first order amplitudes $2\text{Re }\frak{A}_0^*\frak{A}_1$.

\subsubsection{Gaussian pulse}

As a second example we consider a Gaussian weak field~\eqref{WeakGaussianDef}.
The saddle point for the $\omega_1$ integral is given by
\be\label{SigmaGauss1}
\Sigma(\omega_1)=\frac{\sqrt{1+\nu^2+P^2}-i\nu P}{1+\nu^2} \;,
\ee
where $P=p_3/m_\LCperp$ and $\nu=E/\omega^2$. Notice that for this Gaussian pulse the results are conveniently expressed in terms of $\nu$ instead of the usual Keldysh parameter $\gamma$ (at least when considering different orders separately). 
We hence find
\be
\begin{split}
2\text{Im }\varepsilon\Gamma_1=2V_3\int&\frac{\ud^3p}{(2\pi)^3}2\text{Re }\frac{E\varepsilon}{2m_\LCperp\omega}\frac{\sqrt{\pi}}{\Sigma^2}\left[1+\nu^2+\frac{i\nu P}{\Sigma}\right]^{-\frac{1}{2}} \\
&\exp\left\{-\frac{m_\LCperp^2}{E}\left(\frac{\pi}{2}+iP\Sigma+\arccos\Sigma\right)\right\} \;,
\end{split}
\ee
where $\Sigma$ is given by~\eqref{SigmaGauss1}. This is again exactly the same as our result for $2\text{Re }\frak{A}_0^*\frak{A}_1$ in~\cite{Torgrimsson:2017pzs} where we used a WKB approach. This follows immediately from the expressions for the zeroth~\eqref{zerothOrderA} and first order amplitudes~\cite{Torgrimsson:2017pzs}
\be\label{A1Gauss}
\begin{split}
\varepsilon\frak{A}_1=&\delta_{s,s'}\frac{E\varepsilon\sqrt{\pi}}{2m_\LCperp\omega}\frac{1}{\Sigma^2}\left[1+\nu^2+\frac{iP\nu}{\Sigma}\right]^{-\frac{1}{2}} \\
&\exp\left\{-\frac{m_\LCperp^2}{E}\left[iP\Sigma+\arccos\Sigma-i\phi(P)\right]\right\} \;.
\end{split}
\ee

\subsection{Second order $\Gamma_2$}

\begin{figure}
\includegraphics[width=\linewidth]{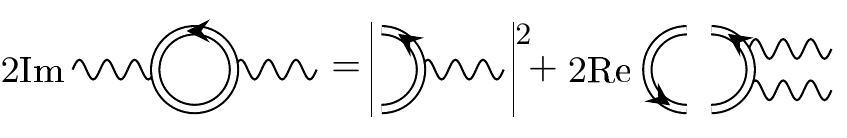}
\caption{A diagrammatic illustration of the relation between the effective action and the amplitude at second order.}
\label{Gamma2AmpAmp-fig}
\end{figure}

At second order there are two different contributions, which in the WKB approach are given by the square of the first order amplitude $|\frak{A}_1|^2$ and the cross term between the zeroth and second order amplitudes $2\text{Re }\frak{A}_0^*\frak{A}_2$. As we will see, we can obtain both of these contributions with the worldline-momentum approach, cf. Fig.~\ref{Gamma2AmpAmp-fig}. 
By expanding~\eqref{worldline-spectrum-start}
to second order we find
\be\label{Gamma2Start}
\begin{split}
\varepsilon^2\Gamma_2=&-V_3\int\frac{\ud^3p}{(2\pi)^3}\int_0^\infty\frac{\ud s}{s}\cot s\; e^{-\frac{s m_\LCperp^2}{E}}\int_0^1\!\ud\tau_1\ud\tau_2 \\
&\int\frac{\ud\omega_1}{2\pi}\frac{\ud\omega_2}{2\pi}
a(\omega_1)a(\omega_2)\Big\{\!\Big[\frac{s}{E}\Big]^2\Big[\frac{\omega_1\omega_2}{2}-2E^2t_1t_2 \\
&-\tan s(\omega_1 Et_2+\omega_2Et_1)\Big]-\frac{s}{E}\delta(\tau_1-\tau_2)\Big\} \\
&e^{\!-\frac{1}{E}\big[ip_3[\omega_1+\omega_2]+\frac{\omega_1^2+\omega_2^2}{4}\cot s+\frac{\omega_1\omega_2}{2}\frac{\cos s(1-2|\tau_1-\tau_2|)}{\sin s}\big]} \;,
\end{split}
\ee
where
$t_1=\frac{1}{2E\sin s}(\omega_1\cos s+\omega_2\cos[s(1-2|\tau_1-\tau_2|)])$
and $t_2=t_1(\omega_1\leftrightarrow\omega_2)$.
We divide $\Gamma_2$ into two parts, one where the two Fourier frequencies have opposite signs and the other where they have the same sign, which we treat separately.

We begin with the region where $\omega_1\omega_2<0$, which gives the dominant contribution. 
Because of the translation symmetry/periodicity in $\tau$ (see e.g.~\cite{Schubert:2001he}), the integrand becomes independent of $\tau_2$ after changing variables from $\tau_1$ to $\tau_1'=\tau_1-\tau_2$. We perform the remaining $\tau_1'$-integral by expanding around the saddle point $\tau_1'=1/2$. 
Next we perform the $s$-integral, for which 
the exponential part of the integrand is given by
\be\label{sExponentwwNegative}
\exp\left\{-\frac{m_\LCperp^2}{E}\left(s+[r_1^2+r_2^2]\cot s+2r_1r_2\csc s\right)\right\} \;,
\ee
where $r_i=\omega_i/2m_\LCperp$.
The saddle point is given by
\be\label{sSaddleGamma2} 
s=2\arccos\sqrt{\frac{1}{2}\left[1-r_1r_2-\sqrt{(1-r_1^2)(1-r_2^2)}\right]}\;,
\ee
where the sign in front of the square root has been determined by demanding that the integral around the saddle point gives a factor of $i$ (as only such a saddle point contributes to $\text{Im }\Gamma$). 
At the saddle point we find\footnote{One can show this e.g by studying the derivative of the exponent with respect to $r_1$ and $r_2$.}
\be
\begin{split}
&\eqref{sExponentwwNegative}\to\exp\bigg\{\!-\frac{m_\LCperp^2}{E}\bigg(\pi-\text{sign}(r_1-r_2) \\
&\left[\arcsin r_1+r_1\sqrt{1-r_1^2}-\arcsin r_2-r_2\sqrt{1-r_2^2}\right]\bigg)\bigg\} \;.
\end{split}
\ee
We have assumed that $\omega_1\omega_2<0$. Without loss of generality we consider $\omega_2<0$ and multiply with a factor of $2$ to account for the other case. Changing variable $\omega_2\to-\omega_2$, this contribution to the second order becomes
\be\label{finalGeneralGamma2}
\begin{split}
2\text{Im }&\varepsilon^2\Gamma_2(\omega_1\omega_2<0)
=2V_3\int\frac{\ud^3p}{(2\pi)^3}\frac{\pi}{E}\bigg|\int_0^\infty\frac{\ud\omega_1}{2\pi} \\
&\frac{a(\omega_1)}{\sqrt{r_1}[1-r_1^2]^\frac{1}{4}}e^{-\frac{m_\LCperp^2}{E}\left(2iPr_1+\arccos r_1-r_1\sqrt{1-r_1^2}\right)}\bigg|^2 \;,
\end{split}
\ee   
where $P=p_3/m_\LCperp$.  
It is now clear that~\eqref{finalGeneralGamma2} agrees with $|\varepsilon\frak{A}_1|^2$, i.e.
\be
2\text{Im }\varepsilon^2\Gamma_2(\omega_1\omega_2<0)=V_3\int\frac{\ud^3p}{(2\pi)^3}\sum_\text{spin}|\varepsilon\frak{A}_1|^2 \;, 
\ee 
where $\Gamma_2(\omega_1\omega_2<0)$ and $\frak{A}_1$ are given by~\eqref{finalGeneralGamma2} and~\eqref{firstOrderA}, respectively.

Next we consider the second region, where $\omega_1\omega_2>0$. For the term without $\delta(\tau_1-\tau_2)$ we use translation invariance to set $\tau_2=1/2$. The exponent is maximized at $\tau_1=1/2$. 
For $\omega_1\omega_2<0$ we could neglect the term with $\delta(\tau_1-\tau_2)$, but this time we need it as it leads to the same exponential as the other terms. 
The exponential for the $s$-integral becomes
$\exp\left\{-\frac{m_\LCperp^2}{E}\left(s+\Sigma^2\cot s\right)\right\}$, where $\Sigma=\frac{|\omega_1+\omega_2|}{2m_\LCperp}$.
This is the same exponential as in~\eqref{sIntGamma1} for the first order, except that $\Sigma$ is now given by the sum of two Fourier frequencies. The saddle point and the integral around it are therefore given by~\eqref{sSaddleGamma1} and~\eqref{sSaddleIntegralGamma1}. The contribution from $\omega_1,\omega_2<0$ is the complex conjugate of that from $\omega_1,\omega_2>0$, and hence
\be\label{Gamma2w1w2SameSign}
\begin{split}
2&\text{Im }\varepsilon^2\Gamma_2(\omega_1\omega_2>0)=-2V_3\int\frac{\ud^3p}{(2\pi)^3}2\text{Re} \\
&\int_0^\infty\frac{\ud\omega_1}{2\pi}\frac{\ud\omega_2}{2\pi}a(\omega_1)a(\omega_2)\sqrt{\frac{\pi}{E}}\frac{2m_\LCperp}{\omega_1\omega_2}\frac{(1-\Sigma^2)^\frac{1}{4}}{\sqrt{\Sigma}} \\
&\exp\left\{-\frac{m_\LCperp^2}{E}\left(\frac{\pi}{2}+2iP\Sigma+\arccos\Sigma-\Sigma\sqrt{1-\Sigma^2}\right)\right\} \;.
\end{split}
\ee
Given the first order result, this looks like it could be the cross term between the zeroth and second order amplitudes $2\text{Re }\frak{A}_0^*\frak{A}_2$. To show that this is indeed the case, we first have to obtain $\frak{A}_2$, which we do in the next section.

Although $\Sigma$ is here given by the sum of two Fourier frequencies, for Sauter-like fields~\eqref{Gamma2w1w2SameSign} still leads to the same exponential as in~\eqref{Gamma1SauterFin} for $\Gamma_1$, and then there is nothing to compensate for the extra factor of the weak field strength $a\sim\varepsilon\ll1$, which means that this second order contribution~\eqref{Gamma2w1w2SameSign} can be neglected. This is why we in~\cite{Torgrimsson:2017pzs} did not have to calculate $\frak{A}_2$ in order to find good agreement with exact/numerical results for Sauter-like fields. As we showed in~\cite{Torgrimsson:2017pzs}, though, for e.g. Gaussian pulses, higher orders can be important.

\section{Using the propagator in a constant electric field}
\label{Using the propagator in a constant electric field}

In this section we show how to extend the WKB approach in~\cite{Torgrimsson:2017pzs} to obtain the amplitude at higher orders. To do so, we use the fermion propagator in a constant electric field. 
The propagator is defined by\footnote{See~\cite{Fradkin:1991zq} for a detailed discussion of different types of propagators.} 
\be
\frac{\langle0,\text{out}|T\Psi_\alpha(x)\bar{\Psi}_\beta(x')|0,\text{in}\rangle}{\langle0,\text{out}|0,\text{in}\rangle}=:iG_{\alpha\beta}(x,x') 
\ee
and satisfies
\be
(i\slashed{\mathcal{D}}_x-m)G(x,x')=\delta(x-x') \;,
\ee
where $\mathcal{D}_\mu=\partial_\mu+i A_\mu$. 
The propagator can be obtained from e.g.~\cite{Schwinger:1951nm,Fradkin:1991zq,Dittrich:2000zu}
\be\label{PropagatorInConstantE}
\begin{split}
G(x,x')=-&e^{-\frac{iE}{2}(z-z')(t+t')}\int\frac{\ud^4q}{(2\pi)^4}e^{-iq(x-x')} \int_0^\infty
\ud s \\
&\exp\left\{-sm_\LCperp^2+(q_0^2-q_3^2)\frac{\tan(Es)}{E}\right\} \\
&\Big[\slashed{q}+m+i(\gamma^0q_3+\gamma^3q_0)\tan(Es)\Big] \\
&\Big[1-i\gamma^0\gamma^3\tan(Es)\Big] \;.
\end{split}
\ee
With the standard $i\epsilon$-prescription $m^2\to m^2-i\epsilon$, the contour for the $s$-integral can be taken along the imaginary axis from $s=0$ to $s=i\infty$ or rotated toward the real axis, but not all the way since there are singularities there due to $\tan s$.

\subsection{Second order $\frak{A}_2$}

\begin{figure}
\includegraphics[width=\linewidth]{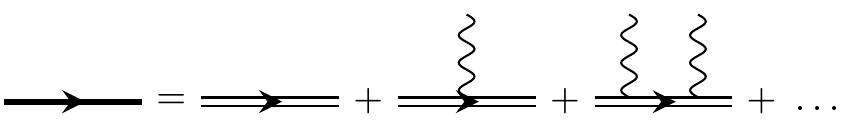}
\caption{The expansion of the pair production amplitude in terms of the weak field, with the same notation as in Fig.~\ref{GammaExpansion1-fig}.}
\label{openFermionLine-fig}
\end{figure}

The second order amplitude is given by (note that $\langle0,\text{out}|0,\text{in}\rangle\approx1$)
\be\label{A2start}
\begin{split}
(2\pi)^3\delta^3({\bf p}+{\bf p'})\varepsilon^2\frak{A}_2=&
(-i)^2\int\ud^4x\ud^4x'\bar{u}_{s,{\bf p}}(t)e^{ip_jx^j} \\
&\slashed{a}(t)iG(x,x')\slashed{a}(t')v_{s',{\bf p'}}(t')e^{ip'_jx'^j} \;.
\end{split}
\ee
This second-order part of the amplitude is represented by the last diagram in Fig.~\ref{openFermionLine-fig}. 
We begin with the trivial spatial integrals, which give the momentum conservation delta function and a second delta function that we use to perform three of the Fourier integrals in the propagator, in particular $q_3=p_3-E(t+t')/2$.
The last term comes from the holonomy factor in the propagator. The reason we cannot neglect this term for $E\ll1$ is that the saddle points for the time integrals turn out to be on the order of $t\sim1/E$.

Next we turn to the proper-time $s$ integral. In the previous sections we used the saddle-point method to perform proper-time integrals in order to obtain the imaginary part of the effective action. For the propagator considered here, though, both its real and imaginary part contribute to the amplitude and  the dominant contribution comes from $s\approx0$. Upon expanding to lowest order in $s$ one finds that the field-dependent propagator reduces to the free propagator times the holonomy factor. 
This means that the factors from the last exponential in~\eqref{PropagatorInConstantE} do not affect the saddle points for the $t$, $t'$ and $q_0$-integrals, they only affect the prefactor.
So, to a first approximation the propagator only gives a field-dependent contribution via the holonomy factor. This approximation leads to results that agree with the approximations we obtain with the worldline formalism in Sec.~\ref{Momentum spectrum from the worldline formalism}, \ref{HigherOrderPreWorldline} and~\ref{AnFromGammanSec}, where $s=0$ corresponds to $\tau_k=\tau_l$ for the $\tau$ variables that correspond to $\omega_k\omega_l>0$, see also~\cite{Torgrimsson:2017pzs,Satunin:2018rdw}. 

We approximate the exact wave functions with the WKB approximations $u\to U$ and $v\to V$ as in~\cite{Torgrimsson:2017pzs} (see Appendix~\ref{Ingredients for the WKB approach}), which leads to the following exponent for the time integrals
\be
\exp\left\{i\int_0^t\!\pi_0-i\omega_1 t-iq_0(t-t')-i\omega_2 t'+i\int_0^{t'}\!\pi_0\right\} \;.
\ee
We perform the integrals over $t$, $t'$ and $q_0$ with the saddle-point method. The saddle point is determined by $\pi_0(t)-\omega_1-q_0=0$, $\pi(t')-\omega_2+q_0=0$ and $t-t'=0$, which give $Et=Et'=p_3+im_\LCperp\sqrt{1-\Sigma^2}$, where $\Sigma=(\omega_1+\omega_2)/(2m_\LCperp)$, and $q_0=(\omega_2-\omega_1)/2$. 
To lowest order in $E$ the proper-time integral simply gives
$\int_0^\infty\ud s\; e^{-\omega_1\omega_2 s}=\frac{1}{\omega_1\omega_2}$.
Since most of this integral comes from the region with $s\lesssim 1/(\omega_1\omega_2)$, we see that our approximation $Es\ll1$ requires $E/(\omega_1\omega_2)\ll1$. For e.g. a Gaussian or a Sauter pulse, $a'(t)\sim e^{-(\omega t)^2}$ or $\text{sech}^2\omega t$, the Fourier integrals are dominated by high-frequency components ($\omega_i\gg\omega$ with $\omega\ll1$) with the saddle points on the order of $\omega_i\sim 1$, which agrees with $E/(\omega_1\omega_2)\ll1$ as $E\ll1$. For a monochromatic field $\sim\cos\omega t$ we only have photons with frequency $\omega$ and then one might want to keep $\omega\ll1$ for experimental reasons. However, one is nevertheless forced to consider larger $\omega$ in the monochromatic case if one wants significant dynamical assistance comparable to the Gaussian or Sauter cases.   
So, for frequencies that give significant enhancement this should be a good first approximation.
 
The final piece comes from the spinor structure in the prefactor, which we calculate using the spinor representation in~\cite{Torgrimsson:2017pzs}. This leads to
$\bar{U}_{s,p}\gamma^3(\slashed{q}+m)\gamma^3V_{s',-p}\to-\delta_{s,s'}2m_\LCperp\frac{\pi_3}{\pi_0}$. 
Collecting all the terms we finally find
\be\label{finalA2general}
\begin{split}
\varepsilon^2\frak{A}_2=&-\delta_{s,s'}\int_0^\infty\frac{\ud\omega_1}{2\pi}\frac{\ud\omega_2}{2\pi}a(\omega_1)a(\omega_2) \\
&\frac{2m_\LCperp}{\omega_1\omega_2}\left[\frac{\pi}{E}\frac{\sqrt{1-\Sigma^2}}{\Sigma}\right]^\frac{1}{2}\exp\left\{\frac{im_\LCperp^2}{E}\phi(P)\right\} \\
&\exp\left\{-\frac{m_\LCperp^2}{E}\left(2iP\Sigma+\arccos\Sigma-\Sigma\sqrt{1-\Sigma^2}\right)\right\} \;,
\end{split}
\ee
where $\Sigma=(\omega_1+\omega_2)/(2m_\LCperp)$.
With the zeroth order amplitude given by~\eqref{zerothOrderA} (note that it contains the same irrelevant phase as in~\eqref{finalA2general})
we immediately see that the cross term between the zeroth and second order amplitudes gives exactly~\eqref{Gamma2w1w2SameSign}, i.e.
\be
2\text{Im }\varepsilon^2\Gamma_2(\omega_1\omega_2>0)=V_3\int\frac{\ud^3p}{(2\pi)^3}\sum_\text{spin}2\text{Re }\frak{A}_0^*\varepsilon^2\frak{A}_2 \;,
\ee
where $\Gamma_2(\omega_1\omega_2>0)$, $\frak{A}_0$ and $\varepsilon^2\frak{A}_2$ are given by~\eqref{Gamma2w1w2SameSign}, \eqref{zerothOrderA} and~\eqref{finalA2general}, respectively, 
and where the sum over spin simply gives a factor of $2$.

In fact, having obtained the second order amplitude, we can now use it to calculate also the prefactor of the dominant contribution to $P_3$ and $P_4$ (from $2\text{Re }\frak{A}_1^*\frak{A}_2$ and $|\frak{A}_2|^2$, respectively).

\subsection{Second order $\frak{A}_2$ for a Gaussian pulse}

Since the first orders dominate for Sauter-like pulses, we turn directly to a Gaussian pulse, for which the dominant contribution can come from higher orders. To perform the Fourier integrals in~\eqref{finalA2general}, we change variables to $\Sigma=(\omega_1+\omega_2)/(2m_\LCperp)$ and $\theta=(\omega_1-\omega_2)/(2m_\LCperp)$ and perform the integrals with the saddle-point method. 
The saddle point is given by $\theta=0$ and $\Sigma=\Sigma_2$, where 
\be\label{SigmanDef}
\Sigma_n=\frac{\sqrt{1+\nu_n^2+P^2}-i\nu_nP}{1+\nu_n^2} \;,
\ee
$\nu_n:=\nu/n$ and $\nu=E/\omega^2$ (these definitions of $\nu_n$ and $\Sigma_n$ also apply to higher orders).
The $\Sigma$ integral is formally the same as in the first-order case~\eqref{SigmaGauss1} after replacing $\nu$ with $\nu_2$. 
Thus, the second-order amplitude for a Gaussian pulse is given by
\be\label{A2Gauss}
\begin{split}
\varepsilon^2\frak{A}_2=&\delta_{s,s'}\!\left[\frac{E\varepsilon}{\omega}\right]^2\!\frac{\sqrt{\pi E\nu_2}}{m_\LCperp^3\Sigma_2^4}\frac{1+\frac{iP}{\nu_2\Sigma_2}}{\left[1+\nu_2^2+\frac{i\nu_2P}{\Sigma_2}\right]^{\frac{1}{2}}} \\
&\exp\left\{-\frac{m_\LCperp^2}{E}\left[iP\Sigma_2+\arccos\Sigma_2-i\phi(P)\right]\right\} \;.
\end{split}
\ee

\begin{figure}
\includegraphics[width=\linewidth]{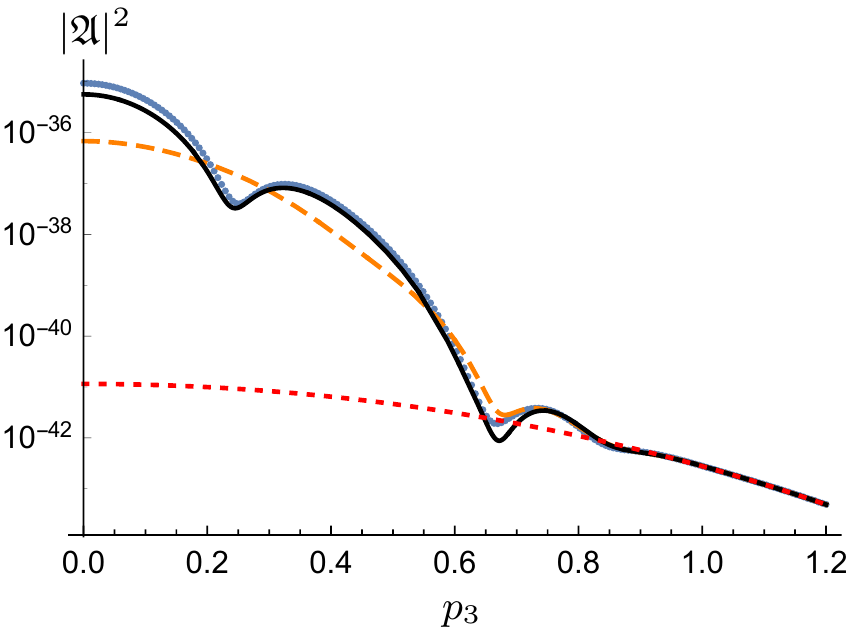}
\caption{The momentum spectrum as a function of the longitudinal momentum $p_3/m$ for $p_\LCperp=0$. The field parameters are chosen as in the right plot of Fig.~3 in~\cite{Torgrimsson:2017pzs}, i.e. $E=0.033E_{\rm crit}$, $\varepsilon=10^{-3}$, $\gamma=3.8$ and, for the numerical results, the strong field is a slowly varying Sauter pulse with $\gamma_{\rm strong}=0.2$. 
The red dotted curve corresponds to the strong field alone 
$|\frak{A}_0|^2$, the orange dashed curve is given by $|\frak{A}_0+\frak{A}_1|^2$, and the black solid curve is given by $|\frak{A}_0+\frak{A}_1+\frak{A}_2|^2$. 
The blue dots show the result obtained in~\cite{Torgrimsson:2017pzs} by numerically solving the Riccati equation using the code from~\cite{Schneider:2016vrl}. It is obvious that for these parameter values we need the second-order amplitude to obtain a good approximation of the probability. (For the weak field alone the spectrum at $p_3=0$ is $\sim10^{-43}$ and quickly becomes much smaller for larger $p_3$.)}
\label{GaussSpecA2-fig}
\end{figure}

In Fig.~\ref{GaussSpecA2-fig} we return to an example which we in~\cite{Torgrimsson:2017pzs} used to demonstrate that $|\frak{A}_0+\frak{A}_1|^2$ is not always enough to obtain a good approximation of the spectrum for these fields. Fig.~\ref{GaussSpecA2-fig} shows that $|\frak{A}_0+\frak{A}_1+\frak{A}_2|^2$, on the other hand, does lead to good agreement with the exact/numerical solution of the Riccati equation that was obtained in~\cite{Torgrimsson:2017pzs}, especially given that the parameter values in this example have not been optimized but are simply the ones we considered in~\cite{Torgrimsson:2017pzs}, and the strong field is actually not a constant field but a slowly varying Sauter pulse. 
As mentioned, the dominant order is given by~\eqref{domNgauss}, which reaches its maximum at $\chi=\sqrt{2}$. For $\varepsilon=10^{-3}$ this corresponds to $\gamma=3.72$, which is close to the value we have chosen in Fig.~\ref{GaussSpecA2-fig}. For this example~\eqref{domNgauss} gives $N_{\rm dom}\sim4$, which agrees with the fact that we need $\mathfrak{A}_2$ to find a good agreement. We can increase the dominant order by decreasing $E$, but this also makes the probability much smaller. 

Together with the first order amplitude~\eqref{A1Gauss} and with the saddle point for the longitudinal momentum given by~\eqref{Psaddlenm} we find that the total probability at third order is given by
\be\label{totalP3}
\begin{split}
\varepsilon^3P_3&=V_3\int\frac{\ud^3p}{(2\pi)^3}\sum_\text{spin}2\text{Re }\varepsilon\frak{A}_1^*\varepsilon^2\frak{A}_2 \\
&=V_3\frac{(E\varepsilon)^3}{6}\sqrt{\frac{E}{\pi}}\frac{\bar{\nu}^\frac{3}{2}(1+\bar{\nu}^2)^\frac{5}{2}}{\arctan\bar{\nu}}\exp\left\{-\frac{2}{E}\arctan\bar{\nu}\right\} \;,
\end{split}
\ee
where $\bar{\nu}=\frac{2N}{N^2-1}\nu=3\nu/4=(\nu_1+\nu_2)/2$. 
From the square of the second-order amplitude we obtain
\be\label{totalP4}
\begin{split}
\varepsilon^4P_4&=V_3\int\frac{\ud^3p}{(2\pi)^3}\sum_\text{spin}|\varepsilon^2\frak{A}_2|^2 \\ &=V_3(E\varepsilon)^4\sqrt{\frac{E}{2\pi}}\frac{\nu_{\scriptscriptstyle2}^\frac{5}{2}(1+\nu_{\scriptscriptstyle2}^2)^\frac{7}{2}}{2\arctan\nu_{\scriptscriptstyle2}}\exp\left\{-\frac{2}{E}\arctan\nu_{\scriptscriptstyle2}\right\} \;.
\end{split}
\ee
Compare~\eqref{totalP3} and~\eqref{totalP4} with Eq.~(2.18) in~\cite{Torgrimsson:2017pzs} for the second-order term, which can be expressed as
\be\label{totalP2}
\varepsilon^2P_2=V_3\frac{(E\varepsilon)^2}{32}\sqrt{\frac{E\nu}{2\pi}}\frac{(1+\nu^2)^\frac{3}{2}}{\arctan\nu}
\exp\left\{\!-\frac{2}{E}\arctan\nu\right\} \;.
\ee  
Both~\eqref{totalP3} and~\eqref{totalP4} are in perfect agreement with $2\text{Im }\Gamma_3$ and $2\text{Im }\Gamma_4$, respectively, which we show using the worldline formalism (not in the momentum representation) in Sec.~\ref{HigherOrderPreWorldline}.

Recall that to obtain the zeroth order, $P_0\sim e^{-\pi/E}$, from a perturbative series, it is necessary to use Borel resummation techniques~\cite{Chadha:1977my,Dunne:1999uy}. The saddle-point results~\eqref{totalP3}, \eqref{totalP4} and \eqref{totalP2} can be expanded in a Taylor series in $E$ (by keeping $\omega$ in $\nu=E/\omega^2$ fixed) and then directly reconstructed without using e.g. Borel resummation. However, this does not mean that one can obtain~\eqref{totalP3}, \eqref{totalP4} and \eqref{totalP2} in the region $\nu\sim1$ from an ab initio perturbative treatment of the strong field.

\subsection{Higher orders $\frak{A}_n$}
\label{Higher orders using the propagator}

We will now use the propagator from the previous section to obtain the exponentials of higher order amplitudes. We obtain the $n$-th order amplitude from
\be
\begin{split}
(2\pi)^3\delta^3({\bf p}+{\bf p'})&\varepsilon^n\frak{A}_n\sim
\int\ud^4x_1...\ud^4x_n
\bar{u}(t_1)e^{ip_jx_1^j} \\
&\slashed{a}(t_1)G(x_1,x_2)\slashed{a}(t_2)G(x_2,x_3)\dots \\
&\slashed{a}(t_{n-1})G(x_{n-1},x_n)\slashed{a}(t_n)v(t_n)e^{ip'_jx_n^j} \;.
\end{split}
\ee
The spatial integrals give delta functions 
which we use to perform the integrals over ${\bf q}^{(j)}$ for each propagator. 
The proper-time integrals from the propagators are again dominated by $s_k\sim0$ and do not affect the exponential behavior of the probability, which means that, when performing the time integrals with the saddle-point method, the exponential is a relatively simple generalization of the second order case above. Using~\eqref{intpi0tophi} and shifting the time variables, $t_k\to t_k+p_3/E$, to make the simple $p_3$-dependence manifest, we find
\be\label{Andwdtdq0shift}
\begin{split}
\varepsilon^n\frak{A}_n\sim\int&\prod_{k=1}^n[\ud\omega_k\ud t_k a(\omega_k)]\prod_{k=1}^{n-1}\ud q^{(k)}_0\dots \\
\exp i&\left\{-\frac{p_3}{E}\sum_{k=1}^n\omega_k+\frac{m_\LCperp^2}{2E}\phi\left[\frac{Et_1}{m_\LCperp}\right]-\sum_{k=1}^n\omega_k t_k \right. \\
&\left.-\sum_{k=1}^{n-1}q^{(k)}_0(t_k-t_{k+1})+\frac{m_\LCperp^2}{2E}\phi\left[\frac{Et_n}{m_\LCperp}\right]\right\} \;,
\end{split}
\ee
where the ellipses stand for factors that do not affect the exponential behavior of the probability
(and we have omitted the term in~\eqref{intpi0tophi} with $\phi(p_3/m_\LCperp)$ since it anyway cancels when squaring the amplitude). 
We perform the $t_1$ integral with the saddle-point method, where the saddle point is given by $Et_1(q^{(1)}_0)=i\sqrt{m_\LCperp^2-(\omega_1+q_0^{(1)})^2}$ (assuming $0<\omega_1+q_0^{(1)}<m_\LCperp$).
We can now perform the $q^{(1)}_0$ integral also with the saddle-point method. Although $t_1(q^{(1)}_0)$ now depends on $q^{(1)}_0$, the saddle-point equation for $q^{(1)}_0$ is simply given by $t_1(q_0^{(1)})=t_2$, and we do not even have to find the explicit solution for $q_0^{(1)}$ in order to obtain the exponential part of the probability.
We can now perform the integrals over $t_2$ and $q^{(2)}_0$ in exactly the same way, the only difference is $\omega_1\to\omega_1+\omega_2$. This in turn leads to similar integrals for $t_3$ and $q^{(3)}_0$, with $\omega_1\to\omega_1+\omega_2+\omega_3$, and so on. 
The last time integral is similar to the previous ones, and the saddle point is given by
$Et_n=im_\LCperp\sqrt{1-\Sigma^2}$, where
$\Sigma=\frac{1}{2m_\LCperp}\sum_{k=1}^n\omega_k$.
The sum over Fourier frequencies is the only difference between the resulting exponent and the one for $n=1$. We can therefore immediately write down the result for arbitrary $n$ using the first order results in~\cite{Torgrimsson:2017pzs}. We hence find
\be\label{AnExpSigma}
\varepsilon^n\frak{A}_n\sim\!\int\!\prod_{k=1}^n\ud\omega_ka(\omega_k)\dots e^{-\frac{m_\LCperp^2}{E}\left[2iP\Sigma+\arccos\Sigma-\Sigma\sqrt{1-\Sigma^2}\right]} \;, 
\ee
where $P=p_3/m_\LCperp$
\be\label{AnExpSigmaSigma}
\Sigma=\frac{1}{2m_\LCperp}\sum_{k=1}^n\omega_k \;,
\ee
and the ellipses stand for factors that do not affect the exponential.

In fact, this exponential part of the amplitude can also be obtained from the worldline-momentum approach: The $n$-th order of the imaginary part of the effective action, $\text{Im }\Gamma_n$, corresponds to the sum of products of different orders of the amplitude. For example, $\text{Im }\Gamma_4$ contains $|\frak{A}_2|^2$, $\text{Re }\frak{A}_1^*\frak{A}_3$ and $\text{Re }\frak{A}_0^*\frak{A}_4$. The $n$-th order amplitude $\frak{A}_n$ can be obtained from the term in $\text{Im }\Gamma_n$ in which all Fourier frequencies have the same sign, because this corresponds to the cross term $2\text{Re }\frak{A}_0^*\frak{A}_n$ and $\frak{A}_0$ has a simple exponential that is easy to separate out. In this case the exponential is given by
\be
e^{-\frac{m_\LCperp^2}{E}\left(2iP\Sigma+s+\frac{1}{4m_\LCperp^2}\sum_{i,j=1}^n\omega_i\omega_j\frac{\cos[s(1-2|\tau_i-\tau_j|)]}{\sin s}\right)} \;,
\ee
which for $\omega_i\omega_j>0$
is maximized by $|\tau_i-\tau_j|=0,1$, which leads to
\be\label{expPsSigmaCot}
\exp\left\{-\frac{m_\LCperp^2}{E}\left(2iP\Sigma+s+\Sigma^2\cot s\right)\right\}
\ee
with the same $\Sigma$ as in~\eqref{AnExpSigmaSigma}. Performing the $s$~integral with the saddle-point method as in~\eqref{sSaddleIntegralGamma1} gives the same exponential for $\frak{A}_n$ as in~\eqref{AnExpSigma}. See Appendix~\ref{AnFromGammanSec} for more details on this approach. 

Upon squaring the amplitude, the $N$-th order terms in the probability are given by $\frak{A}_{N-n}^*\frak{A}_n$, with $0\leq n\leq N$. Since $p_3$ only enters in the linear term in the exponential, the integral over $p_3$ gives a delta function $\delta(\Sigma'-\Sigma)$, with $\Sigma$ and $\Sigma'$ for $\frak{A}_{N-n}^*$ and $\frak{A}_n$, respectively. This is the same as in Eq.~(5.1) in~\cite{Torgrimsson:2017pzs} and we immediately recover the exponent in Eq.~(5.5) in~\cite{Torgrimsson:2017pzs}, which we there obtained with a completely different approach. Thus, for the total/integrated probability, we can stop at this point; after reproducing Eq.~(5.5) in~\cite{Torgrimsson:2017pzs}, which holds for quite general field shapes of the weak field, the rest of the calculation is identical to that in~\cite{Torgrimsson:2017pzs}. See though Appendix~\ref{higherOrderAnGaussian} for a different approach.

\subsection{Third order $\frak{A}_3$ for a Gaussian pulse}

Having obtained the saddle points at arbitrary orders, it is now straightforward to calculate the prefactors. In this section we do so for the third order amplitude for a Gaussian pulse. The calculation is similar to the one above for $\frak{A}_2$ so we simply state the results. We find
\be\label{A3Gauss}
\begin{split}
\varepsilon^3\frak{A}_3=&\delta_{s,s'}\left[\frac{E\varepsilon}{\omega}\right]^3\frac{27\sqrt{3\pi}E}{128m_\LCperp^5\Sigma_3^8\nu_3}\frac{9-8\Sigma_3^2}{\sqrt{1+\nu_3^2+\frac{i\nu_3P}{\Sigma_3}}} \\
&\exp\left\{-\frac{m_\LCperp^2}{E}\left[iP\Sigma_3+\text{arccos}\Sigma_3-i\phi(P)\right]\right\} \;,
\end{split}
\ee 
where $\Sigma_3$ is given by~\eqref{SigmanDef}. We show in Appendix~\ref{AnFromGammanSec} how to obtain~\eqref{A3Gauss} with the worldline-momentum approach. From~\eqref{A3Gauss} and~\eqref{A2Gauss} we obtain the dominant contribution to $P_5$ and $P_6$,
\be\label{totalP5}
\begin{split}
&\varepsilon^5P_5=V_3\int\frac{\ud^3p}{(2\pi)^3}\sum_\text{spin}2\text{Re }\varepsilon^2\frak{A}_2^*\varepsilon^3\frak{A}_3= \\
&V_3(E\varepsilon)^5\frac{243}{640}\sqrt{\frac{3E}{\pi}}\frac{\bar{\nu}^\frac{3}{2}(1+\bar{\nu}^2)^\frac{9}{2}(1+9\bar{\nu}^2)}{\arctan\bar{\nu}}e^{-\frac{2}{E}\arctan\bar{\nu}} \;,
\end{split}
\ee
where $\bar{\nu}=5\nu/12$, and
\be\label{totalP6}
\begin{split}
&\varepsilon^6P_6=V_3\int\frac{\ud^3p}{(2\pi)^3}\sum_\text{spin}|\varepsilon^3\frak{A}_3|^2= \\ &V_3(E\varepsilon)^6\frac{59049}{131072} \sqrt{\frac{E\nu_{\scriptscriptstyle3}}{2\pi}}\frac{(1+\nu_{\scriptscriptstyle3}^2)^\frac{11}{2}(1+9\nu_{\scriptscriptstyle3}^2)^2}{\arctan\nu_{\scriptscriptstyle3}}e^{-\frac{2}{E}\arctan\nu_{\scriptscriptstyle3}} \;.
\end{split}
\ee

For the example in Fig.~\ref{GaussSpecA2-fig} we can now check that $\frak{A}_3$ indeed gives a negligible contribution to the spectrum, and from~\eqref{totalP2}, \eqref{totalP3}, \eqref{totalP4}, \eqref{totalP5} and~\eqref{totalP6} we find that $\varepsilon^NP_N$ increases from $N=0$ to $N=4$ and then decreases, so for this particular example we do not have to calculate more terms.

\subsection{Cos-Gaussian pulse}

\begin{figure}
\includegraphics[width=\linewidth]{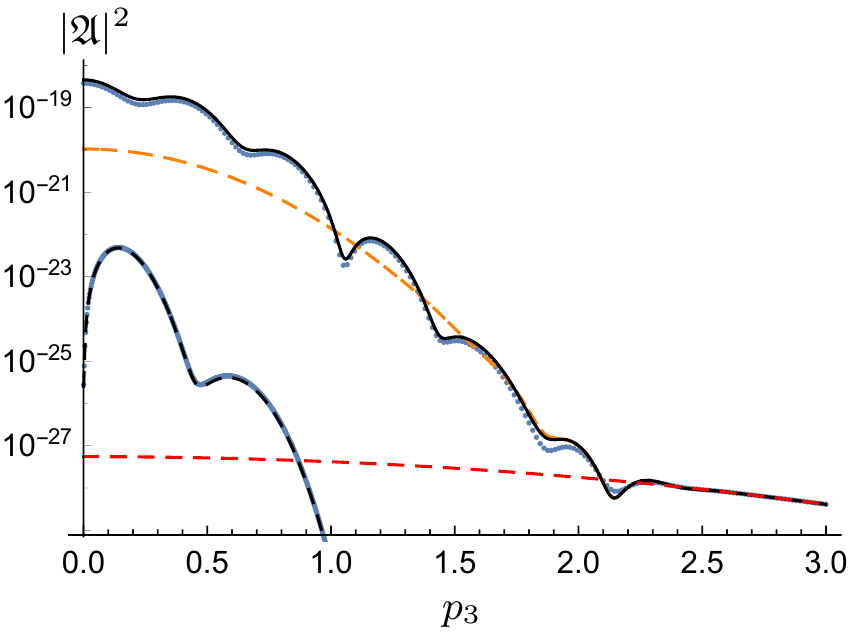}
\caption{The $p_3$ spectrum $|\frak{A}|^2$ at $p_\LCperp=0$ for~\eqref{daCosGauss} with $E=0.05$, $\varepsilon=10^{-3}$, $\omega=1.5E$, $\Omega=0.75$ and $\varphi=0$. The strong field is a Sauter pulse with frequency $E/15$. The red dashed curve gives $|\frak{A}_0|^2$, the orange dashed curve $|\frak{A}_0+\frak{A}_1|^2$ and the black curve $|\frak{A}_0+\frak{A}_1+\frak{A}_2|^2$, where $\frak{A}_1$ and $\frak{A}_2$ are obtained from~\eqref{A1cosGauss} and~\eqref{A2cosGauss}. The blue dotted lines give the exact result obtained by solving the Riccati equation numerically~\cite{Dumlu:2011rr} with the approach in~\cite{Schneider:2016vrl}, i.e. by using the TIDES differential equation solver~\cite{tidesRef} and the multiple-precision library MPFR~\cite{MPFRref}. 
The lower blue and the dashed black curves show the spectrum for the weak field alone, where the dashed black curve is given by $|\frak{A}_1+\frak{A}_2+\frak{A}_3|^2$, with $\frak{A}_i$ from~\eqref{A1cosGaussWeak}, \eqref{A2cosGaussWeak} and~\eqref{A3cosGaussWeak}. This weak spectrum is dominated by $\frak{A}_2$ for $p_3\lesssim0.4$ (except close to $p_3=0$ where $\frak{A}_2=0$) and by $\frak{A}_3$ for $p_3\gtrsim0.5$, while $\frak{A}_1$ is completely negligible.}
\label{cosGaussFig}
\end{figure}

So far we have focused on fields with a single maximum in $t$. However, since it is the Fourier transform of the weak field that is most important here, it is relatively easy to generalize the results in the previous sections to oscillating fields. As an example we consider a sinusoidal field with a Gaussian envelope
\be\label{daCosGauss}
a'(t)=E\varepsilon\cos(\Omega t+\varphi)e^{-(\omega t)^2} \;.
\ee
The Fourier transform is similar to the simple Gaussian pulse,
\be
a(\omega_1)=\frac{\omega_1-\Omega}{2\omega_1}e^{-i\varphi}a_{\rm G}(\omega_1-\Omega)+\frac{\omega_1+\Omega}{2\omega_1}e^{i\varphi}a_{\rm G}(\omega_1+\Omega) \;,
\ee  
where $a_{\rm G}(\omega_1)$ is the Fourier transform for $\Omega=\phi=0$ given by~\eqref{WeakGaussianDef}. If we assume that $\Omega$ is not too small, then one can neglect $a_{\rm G}(\omega_1+\Omega)$ compared to $a_{\rm G}(\omega_1-\Omega)$. We can perform the integrals with the same methods as before, so we simply state the final results here. We find 
\be\label{A1cosGauss}
\begin{split}
\varepsilon\mathfrak{A}_1=&\delta_{s,s'}\frac{e^{-i\varphi}E\varepsilon}{2\omega}\frac{\sqrt{\pi}}{2m_\LCperp\Sigma_1^2} \\
&\frac{e^{-\frac{m_\LCperp^2}{E}[\Lambda_1\nu_1(\Lambda_1-\Sigma_1)+iP\Sigma_1+\text{arccos}\Sigma_1-i\phi(P)]}}{\sqrt{1+\nu_1^2\left(1-\frac{\Lambda_1}{\Sigma_1}\right)+\frac{iP\nu_1}{\Sigma_1}}} \;,
\end{split}
\ee
\be\label{A2cosGauss}
\begin{split}
\varepsilon^2\mathfrak{A}_2=&\delta_{s,s'}\left[\frac{e^{-i\varphi}E\varepsilon}{2\omega}\right]^2\frac{\sqrt{\pi E\nu_2}}{m_\LCperp^3\Sigma_2^4}\left(1-\frac{\Lambda_2}{\Sigma_2}+\frac{iP}{\nu_2\Sigma_2}\right) \\
&\frac{e^{-\frac{m_\LCperp^2}{E}[\Lambda_2\nu_2(\Lambda-\Sigma_2)+iP\Sigma_2+\text{arccos}\Sigma_2-i\phi(P)]}}{\sqrt{1+\nu_2^2\left(1-\frac{\Lambda_2}{\Sigma_2}\right)+\frac{iP\nu_2}{\Sigma_2}}} \;,
\end{split}
\ee
\be\label{A3cosGauss}
\begin{split}
\varepsilon^3\mathfrak{A}_3=&\delta_{s,s'}\left[\frac{e^{-i\varphi}E\varepsilon}{2\omega}\right]^3
\frac{27\sqrt{3\pi}E}{128m_\LCperp^5\Sigma_3^8\nu_3}(9-8\Sigma_3^2) \\
&\frac{e^{-\frac{m_\LCperp^2}{E}[\Lambda_3\nu_3(\Lambda_3-\Sigma_3)+iP\Sigma_3+\text{arccos}\Sigma_3-i\phi(P)]}}{\sqrt{1+\nu_3^2\left(1-\frac{\Lambda_3}{\Sigma_3}\right)+\frac{iP\nu_3}{\Sigma_3}}} \;,
\end{split}
\ee
where $\nu_n=\nu/n$,
\be
\Sigma_n=\frac{\Lambda_n\nu_n^2-iP\nu_n+\sqrt{1+\nu_n^2+P^2-\Lambda_n^2\nu_n^2+2iP\Lambda_n\nu_n}}{1+\nu_n^2}
\ee
and $\Lambda_n=n\Omega/(2m)$.
In Fig.~\ref{cosGaussFig} we compare these terms with the exact numerical result. In this example $|\frak{A}_0+\frak{A}_1|^2$ is not enough, not even qualitatively. 
However by including the second order amplitude, $|\frak{A}_0+\frak{A}_1+\frak{A}_2|^2$, we find a good agreement.

One advantage of this approach is that it gives the correct results in the limits where either the weak or the strong field vanishes. The limit $\varepsilon\to0$ gives trivially the zeroth order $P_0$, which only depends on the strong field. In the other limit we can directly obtain the results by taking $E\to0$ with $E\varepsilon$ fixed in~\eqref{A1cosGauss}, \eqref{A2cosGauss} and~\eqref{A3cosGauss}, which gives
\be\label{A1cosGaussWeak}
\varepsilon\frak{A}_1=\frac{\sqrt{\pi}}{4}\frac{m_\LCperp E\varepsilon}{p_0^2\omega}e^{-i\varphi-\frac{(2p_0-\Omega)^2}{4\omega^2}} \;,
\ee  
\be\label{A2cosGaussWeak}
\varepsilon^2\frak{A}_2=\frac{i}{2}\sqrt{\frac{\pi}{2}}\frac{m_\LCperp p_3(E\varepsilon)^2}{p_0^5\omega}e^{-2i\varphi-\frac{(2p_0-2\Omega)^2}{8\omega^2}} \;,
\ee
\be\label{A3cosGaussWeak}
\varepsilon^3\frak{A}_3=\frac{81\sqrt{3\pi}}{1024}\frac{(9m_\LCperp^2-8p_0^2)m_\LCperp(E\varepsilon)^3}{p_0^8\omega}e^{-3i\varphi-\frac{(2p_0-3\Omega)^2}{12\omega^2}} \;,
\ee
where $p_0=\sqrt{m_\LCperp^2+p_3^2}$. Fig.~\ref{cosGaussFig} shows one example where the dominant contribution comes from $\frak{A}_2$ in one part of the spectrum and from $\frak{A}_3$ in the other, and the agreement with the exact numerical result is excellent. In the limit of a long pulse $\omega\to0$ these terms become proportional to $\delta(2p_0-n\Omega)$ as expected.

\section{Higher-order prefactors for the integrated probability}\label{HigherOrderPreWorldline}

In this section we show how to obtain higher orders of the integrated probability, including the prefactors, using the worldline formalism. We show in particular how to use this method to obtain~\eqref{totalP3}, \eqref{totalP4}, \eqref{totalP2}, \eqref{totalP5} and~\eqref{totalP6}. Our starting point is~\eqref{GammaUsual} with the spin factor given by~\eqref{simpleSpin}. However, as we in this section only calculate the integrated probability, we do not go over to the worldline-momentum representation. This is a generalization of the approach we used in~\cite{Torgrimsson:2017pzs}.
We expand the effective action in the weak field as in~\eqref{Gamma012dots}, where now
\be\label{GammaExpandedGeneral}
\begin{split}
	\varepsilon^N\Gamma_N=&\int\prod_{k=1}^N\frac{\ud\omega_k}{2\pi}a(\omega_k)\int_0^\infty\frac{\ud T}{T}\int_0^1\prod_{k=1}^N\ud\tau_k\oint\mathcal{D}x 
	W_N \\
	&
	\exp\left\{\!-i\left(\frac{T}{2}+\sum_{i=1}^N\omega_i t(\tau_i)+\int_0^1\frac{\dot{x}^2}{2T}+Et\dot{z}\right)\right\} \;,
\end{split}
\ee  
and the prefactor $W_N(T,\omega_i,\dot{z}(\tau_i))$ is obtained from the expansion of 
\be\label{givesWN}
2\cos\left(\frac{iT}{2}\left[E+\int_0^1a'(t)\right]\right)\exp\left(-i\int_0^1 a\dot{z}\right)
\ee
in the field strength $a$. 
We start with the path integral. The transverse integrals simply give
\be
\oint\mathcal{D}x^\LCperp\exp\left(-i\int\frac{-\dot{x}_\LCperp^2}{2T}\right)=\frac{V_\LCperp}{(2\pi iT)^\frac{d}{2}} \;,
\ee
where $d$ is the number of transverse dimensions. 
We separate the time integral into a `center of mass' plus oscillating terms, $t(\tau)\to t_c+t(\tau)$, where the new $t$ obeys $\int_0^1 t=0$. The $t_c$ integral gives a delta function for the Fourier frequencies
$\int\ud t_c\to2\pi\delta(\omega_1+\dots+\omega_N)$.
For the fields we consider here it is natural to switch to Euclidean variables, $t\to-it$ and $T\to-iT$. 
It turns out to be convenient to use $s=ET/2$ instead of $T$. Selecting the $N$-th order from~\eqref{givesWN} 
and exponentiating the resulting products (cf.~\cite{Schubert:2001he}) give
\be
\begin{split}
	W_N=&\text{linear}_\epsilon\frac{i^N}{N!}\left\{\exp\left[is-\sum_{k=1}^N\epsilon_k\left(\dot{z}(\tau_k)+\frac{s}{E}i\omega_k\right)\right]\right. \\
	&\left.
	+\exp\left[-is-\sum_{k=1}^N\epsilon_k\left(\dot{z}(\tau_k)-\frac{s}{E}i\omega_k\right)\right]\right\} \;,
\end{split}
\ee
where $\text{linear}_\epsilon$ selects all the terms that are linear in all $\epsilon_k$\footnote{Note that $\epsilon_k$ ($k=1,\dots,N$) are just temporary, non-physical parameters, which are introduced as a mathematical tool.}. The path integral is now a relatively simple Gaussian. 
We remove the terms in the exponent that are linear in $z$ by making a shift in the integration variables, $z\to z_{\rm cl}+z$, where the ``classical'' part is given by
\be
\dot{z}_{\rm cl}(\tau)=-ET t-T\sum_{k=1}^N\epsilon_k[\delta(\tau-\tau_k)-1] \;.
\ee   
The $z$~integral is now free and gives a volume factor $\Delta z$,
\be
\oint\!\mathcal{D}z\exp\left(-\int\frac{\dot{z}^2}{2T}\right)=\frac{\Delta z}{\sqrt{2\pi T}} \;.
\ee
For the remaining $t$~integral we again make the exponent quadratic by shifting the integration variable, $t\to t_{\rm cl}+t$ where the ``classical'' part is obtained by expanding its equation of motion (cf.~\eqref{tcleom1}),
\be
(\partial_\tau^2+[2s]^2)t_\text{cl}(\tau)=T\sum\limits_{k=1}^N(\omega_k-2s\epsilon_k)(\delta_{\tau,\tau_k}-1) \;,
\ee 
in terms of Fourier modes, which yields (cf.~\eqref{tclsum1})
\be
\begin{split}
	&t_\text{cl}(\tau)=T\sum\limits_{k=1}^N(\omega_k-2s\epsilon_k)\sum_{n\ne0}\frac{e^{2\pi i n(\tau-\tau_k)}}{(2s)^2-(2\pi n)^2} \\
	&=\frac{1}{2E}\sum_{k=1}^N(\omega_k-2s\epsilon_k)\left\{\frac{\cos\big[s(1-2|\tau-\tau_k|)\big]}{\sin s}-\frac{1}{s}\right\} \;,
\end{split}
\ee
where the sum over $n$ can be performed using Eq.~(1.445.2) in~\cite{GradshteynRyzhik}. We perform the Gaussian path integral by Fourier expanding $t$ as in~\eqref{tFourierExpand}
and then multiplying the eigenvalues as in~\eqref{tpathbeta}, which gives
\be
\oint\mathcal{D}t\exp\Big(-\int\frac{t(-\partial_\tau^2-(2s)^2)t}{2T}\Big)=\frac{1}{\sqrt{2\pi T}}\frac{s}{\sin s} \;.
\ee  
The prefactor $W_N$ is now given by
\be\label{WNfromlinear}
\begin{split}
	W_N=&2\frac{i^N}{N!}\text{linear}_\epsilon\cos\left\{s-\frac{s}{E}\sum_{k=1}^N\epsilon_k\omega_k\right\}\exp\left\{\frac{s}{E}\sum_{k=1}^N\epsilon_k\xi_k\right. \\
	&\left.+\frac{s}{E}\sum_{k,l=1}^N\epsilon_k\epsilon_l\left[\delta_{\tau_k\tau_l}-s\frac{\cos[s(1-2|\tau_k-\tau_l|)]}{\sin s}\right]\right\}
	\;,
\end{split}
\ee
where
\be
\xi_k:=\sum_{l=1}^N\omega_l\frac{\cos[s(1-2|\tau_k-\tau_l|)]-\cos s}{\sin s} \;.
\ee 
With the path integral performed, we now have
\be\label{GammaExpandedEuclidean}
\begin{split}
	\varepsilon^N\Gamma_N=&-V_3\int\prod_{k=1}^N\left[\frac{\ud\omega_k}{2\pi}a(\omega_k)\right]2\pi\delta\left(\sum_{k=1}^N\omega_k\right) \\ 
	&\int_0^\infty\!\ud s\left[\frac{E}{4\pi s}\right]^{\frac{d}{2}+1}\frac{1}{\sin s} \int_0^1\prod_{k=1}^N\ud\tau_k\,
	W_N \\
	&e^{\!-\frac{1}{E}\big(s+\sum_{i,j=1}^N\omega_i\omega_j\frac{\cos[s(1-2|\tau_i-\tau_j|)]-\cos s}{4\sin s}\big)} \;.
\end{split}
\ee  
Eq.~\eqref{GammaExpandedEuclidean} complements Eq.~(5.1) in~\cite{Torgrimsson:2017pzs} by providing the prefactor, and so gives the exact $\Gamma_N$ for arbitrary $N$.
In deriving~\eqref{GammaExpandedEuclidean} we have used the fact that $\omega_1+\dots+\omega_N=0$. For $\Gamma_1$, though, one has to be more careful since $a(\omega_1)\omega_1\delta(\omega_1)$ leads to a nonzero contribution. However, as mentioned, we are not interested in terms like $\Gamma_1$ (integrated over the momentum), which have the same exponential as $\Gamma_0$ and therefore only give small corrections.  

We also perform the $\tau_i$ and $s$ integrals for general weak field $a$. Let us first consider the zeroth order as a check of e.g. signs and factors of 2. 
To zeroth order the prefactor is given by $W_0=2\cos s$, the delta function gives a volume factor $2\pi\delta(0)=\Delta t$ and we recover the well-known Euler-Heisenberg action for a constant electric field, see e.g.~\cite{Dunne:2004nc},
\be
\Gamma_0=-V_4\int_0^\infty\!\ud s\left[\frac{E}{4\pi s}\right]^{\frac{d}{2}+1}\frac{2\cos s}{\sin s} e^{-s/E} \;.
\ee
The integration over the first pole gives the leading order of the imaginary part of the effective action as in~\eqref{ImGamma0fin}.

\subsection{$\text{Im }\Gamma_2$}

As a more nontrivial check of~\eqref{GammaExpandedEuclidean}, we compare with previous results for $\Gamma_2$.
In order to compare with the exact expression in~\cite{Dittrich:2000zu} for the polarization tensor in a constant electric field, we make a partial integration in $\tau_1$ to replace the delta function in~\eqref{WNfromlinear}, cf.~\cite{Schubert:2000yt}. 
Using the translation invariance we put $\tau_2=0$ in the integrand. To facilitate comparison with~\cite{Dittrich:2000zu}, we change variable from $\tau_1$ to $v=2\tau_1-1$. We find
\be\label{Gamma2withv}
\begin{split}
	\varepsilon^2\Gamma_2=-V_3&\int\frac{\ud\omega_1}{2\pi}|\omega_1a(\omega_1)|^2\int_0^\infty\!\ud s\left[\frac{1}{4\pi}\right]^{\frac{d}{2}+1}\left[\frac{E}{s}\right]^{\frac{d}{2}-1} \\
	&\int_{-1}^1\ud v 
	\frac{\cos s-\cos(sv)}{\sin^3 s}e^{\!-\frac{1}{E}\left(s-\omega_1^2\frac{\cos(sv)-\cos s}{2\sin s}\right)} \;.
\end{split}
\ee  
For $d=2$, \eqref{Gamma2withv} is identical to the expression we used in~\cite{Torgrimsson:2017pzs} to obtain $P_2$ from the exact polarization tensor in~\cite{Dittrich:2000zu}. 
So we already know that performing the integrals in~\eqref{Gamma2withv} with the saddle-point method leads to a result that agrees with the WKB-based approach we used in~\cite{Torgrimsson:2017pzs}. However, we go through the calculation here to prepare for the calculation of higher-order terms. For a Gaussian weak field, the $\omega_1$~integral is Gaussian and can be performed exactly at this stage. However, since we want to make as much progress as possible for general pulse shapes, we keep the $\omega_1$~integral and perform the other integrals first. 

The saddle point for the $\tau_1$~integral is $\tau_1=1/2$. 
The exponential for the $s$-integral is now given by
\be\label{expaftertauints}
\exp\left\{\!-\frac{2}{E}\left(\frac{s}{2}-\Sigma^2\tan\frac{s}{2}\right)\right\} \;,
\ee
where $\Sigma=|\omega_1|/2$. As performing this proper-time integral with the saddle-point method is similar to what we did for~\eqref{sIntGamma1}, we just state the results here.
The saddle point is given by
\be\label{sSaddleSigma}
s=2\arccos\Sigma \;,
\ee
and the Gaussian integral around it is similar to~\eqref{sSaddleIntegralGamma1}.
After performing all integrals except for the one over $\omega_1$, we find
\be\label{Gamma2onlydwleft}
\begin{split}
	2\text{Im }\varepsilon^2\Gamma_2=V_3&\int\frac{\ud\omega_1}{2\pi}|a(\omega_1)|^2\left[\frac{E}{4\pi s}\right]^\frac{d}{2}\frac{1}{2\Sigma\sqrt{1-\Sigma^2}} \\
	&\exp\bigg\{\!-\frac{2}{E}\bigg(\arccos\Sigma-\Sigma\sqrt{1-\Sigma^2}\bigg)\bigg\} \;,
\end{split}
\ee  
where $s$ is given by~\eqref{sSaddleSigma} and $\Sigma=|\omega_1|/2$. It is now straightforward to check that this agrees with the WKB result: Just square the first order amplitude $\frak{A}_1$, given by~\eqref{firstOrderA}, and integrate over the momenta as in~\eqref{finalGeneralGamma2}. The $p_3$ integral gives a delta function setting the Fourier frequency in $\frak{A}_1^*$ equal to that in $\frak{A}_1$, and the perpendicular momentum integrals are Gaussian around $p_\LCperp\approx0$ and give the $[E/(4\pi s)]^{d/2}$ factor in~\eqref{Gamma2onlydwleft}. 

For a Gaussian field~\eqref{WeakGaussianDef}, we find for the $\omega_1$ integral two saddle points given by $|\omega_1|=2\Sigma$, where
\be\label{Sigmafromnu}
\Sigma=\frac{1}{\sqrt{1+\nu^2}} \;.
\ee
The saddle point~\eqref{Sigmafromnu} 
is relevant also at higher orders, but with $\nu$ depending on the order.
With these two saddle points we find $2\text{Im }\Gamma_2=P_2$ with $P_2$ given by~\eqref{totalP2} (for $d=2$).

\subsection{$\text{Im }\Gamma_3$}

Now we turn to the first nontrivial odd term, $\Gamma_3$, which is illustrated by the fourth diagram  on the right-hand side in Fig.~\ref{GammaExpansion1-fig}.
Because of $\delta(\omega_1+\omega_2+\omega_3)$, one of the three $\omega_i$ must have opposite sign compared to the other two. We assume without loss of generality that $\omega_1$ and $\omega_2$ have the same sign, and multiply with a factor of $3$ to account for the other two equivalent regions. We have two different contributions to $W_3=W^{(1)}_3+W^{(2)}_3$: one ($W^{(1)}_3$) without delta functions,
and the other ($W^{(2)}_3$) with delta functions. 
For $W^{(1)}_3$ we use translation invariance~\cite{Schubert:2001he} to set $\tau_3=0$. Looking at the behavior of the exponential, we find that the dominant contribution comes from the integration region near $\tau_1=\tau_2=1/2$. We expand around this point, $\tau_1=1/2+\delta\tau_1$ and $\tau_2=1/2+\delta\tau_2$. 
We change variable from $\delta\tau_2$ to $\delta\tau_2'=\delta\tau_2-\delta\tau_1$. The leading order perturbation around the ``saddle point'' is given by
\be\label{generalizedSaddle}
\int_{-\infty}^{\infty}\ud\delta\tau_1\ud\delta\tau_2'
\exp\left\{-\frac{1}{E}\left(\frac{(2\Sigma s)^2}{\sin s}\delta\tau_1^2+s\omega_1\omega_2|\delta\tau_2'|\right)\right\} \;.
\ee
From this we see that $\delta\tau_1\sim\sqrt{E}$ while $\delta\tau_2'\sim E$, which means that to leading order we can neglect terms like $\delta\tau_1\delta\tau_2'\sim E^{3/2}$ or $\delta\tau_2'^2\sim E^2$. Note that while the $\delta\tau_1$ integral is Gaussian around the saddle point, the exponent behaves as $|\delta\tau_2'|$ rather than $\delta\tau_2'^2$, so we are dealing here with a generalization of the ordinary saddle-point method. The resulting integrals are still elementary though.
At higher orders we have more terms where the fluctuation, $\delta$ say, around some ``saddle point'' for the $\tau$ integrals behaves as $|\delta|$ rather than $\delta^2$. 
Of the three terms in~$W^{(2)}_3$, we can neglect those with $\delta_{\tau_1,\tau_3}$ and $\delta_{\tau_2,\tau_3}$ since they give exponentially smaller contributions. The term with $\delta_{\tau_1,\tau_2}=\delta(\delta\tau_2')$ leads to the same exponential as the terms in $W_3^{(1)}$. We see from~\eqref{generalizedSaddle} that compared to the integrals in $W_3^{(1)}$ this delta function gives
\be\label{deltaReplace}
\delta(\delta\tau_2')\to\frac{s\omega_1\omega_2}{2E} \;,
\ee
which means that also the prefactor part of this contribution is on the same order as $W_3^{(1)}$. 

The $s$-dependent part of the exponential is now given by~\eqref{expaftertauints} with $\Sigma$ given by 
$\Sigma=|\omega_1+\omega_2|/2$,
and the saddle point is given by~\eqref{sSaddleSigma}. 
The contribution from $\omega_1,\omega_2<0$ is equal to minus the complex conjugate of the contribution from $\omega_1,\omega_2>0$. We hence find
\be\label{Gamma3onlydwlefttot}
\begin{split}
	2\text{Im }\varepsilon^3\Gamma_3=&4V_3\text{Im}\int\frac{\ud\omega_1}{2\pi}\frac{\ud\omega_2}{2\pi}a(\omega_1)a(\omega_2)a(-\omega_1-\omega_2) \\
	&\left[\frac{E}{4\pi s}\right]^\frac{d}{2}\frac{1}{\omega_1\omega_2\Sigma}e^{\!-\frac{2}{E}(\arccos\Sigma-\Sigma\sqrt{1-\Sigma^2})} \;,
\end{split}
\ee  
where $\omega_1>0$, $\omega_2>0$ and $\Sigma=(\omega_1+\omega_2)/2$.
It is now straightforward to check that~\eqref{Gamma3onlydwlefttot} agrees with our WKB results for the amplitude: Just take $\frak{A}_1$ and $\frak{A}_2$ from~\eqref{firstOrderA} and~\eqref{finalA2general}, and integrate $2\text{Re }\frak{A}_1^*\frak{A}_2$ as in~\eqref{totalP3}. The momentum integrals are similar to the previous section and we hence find
\be
2\text{Im }\varepsilon^3\Gamma_3\approx V_3\int\frac{\ud^3p}{(2\pi)^3}\sum_\text{spin}2\text{Re }\varepsilon\frak{A}_1^*\varepsilon^2\frak{A}_2 \;.
\ee    
(Note that we have $\approx$ because the exact relation between the effective action and the amplitude at this order also includes the subleading term with $2\text{Re }\frak{A}_0^*\frak{A}_3$.)

\subsection{$\text{Im }\Gamma_4$}

The effective action at fourth order, $\Gamma_4$, is represented by the fifth diagram on the right-hand side in Fig.~\ref{GammaExpansion1-fig}. 
The dominant contribution to $\Gamma_4$ comes from the region where two $\omega_i$'s are positive and the other two are negative. Without loss of generality we assume $\omega_1,\omega_2>0$ and $\omega_3,\omega_4<0$, and multiply with a factor of $6$ to account for the other equivalent regions. We again use the translation invariance to put $\tau_4=\text{constant}:=\tau_0$, and for definiteness we choose $0<\tau_0<1/2$. Then the dominant contribution comes from the region around $\tau_1=\tau_2=\tau_0+1/2$ and $\tau_3=\tau_0$. Expanding around this point, $\tau_{1,2}=\tau_0+1/2+\delta\tau_{1,2}$ and $\tau_3=\tau_0+\delta\tau_3$, we find two integrals with the $|\delta|$-type of fluctuation and one Gaussian integral,
\be
\exp\left\{-\frac{1}{E}\left(\frac{(2\Sigma s)^2}{\sin s}\delta\tau_1^2+s\omega_1\omega_2|\delta\tau_2'|+s\omega_3\omega_4|\delta\tau_3|\right)\right\} \;,
\ee	
where $\delta\tau_2'=\delta\tau_2-\delta\tau_1$.
The exponential for the $s$~integral has the same form as before, \eqref{expaftertauints}, and hence the saddle point is given by~\eqref{sSaddleSigma}, where $\Sigma=(\omega_1+\omega_2)/2$. 
$W_4$ is given by~\eqref{WNfromlinear} with $\xi_1=\xi_2=-\xi_3=-\xi_4=-2\sqrt{1-\Sigma^2}$, We can calculate the delta function terms in $W_4$ by reexpressing the delta functions using partial integration, but it is easier to use the delta functions to perform $\tau$-integrals. We first note that, to leading order in $E$, we can take $\epsilon_k\epsilon_l[\delta_{\tau_k,\tau_l}...]\to\epsilon_k\epsilon_l\delta_{\tau_k,\tau_l}$ in~\eqref{WNfromlinear}, and we only need to consider the terms with $\delta_{\tau_1,\tau_2}$ and $\delta_{\tau_3,\tau_4}$, which contribute similarly to~\eqref{deltaReplace}, since the other delta functions lead to exponentially smaller contributions. 
We hence find
\be\label{Gamma4onlydwleft}
\begin{split}
	2\text{Im }\varepsilon^4\Gamma_4=&V_3\int\prod_{k=1}^4\left[\frac{\ud\omega_k}{2\pi}a(\omega_k)\right]2\pi\delta\left(\sum_{k=1}^4\omega_k\right)\left[\frac{E}{4\pi s}\right]^\frac{d}{2}  \\
	&\frac{4\sqrt{1-\Sigma^2}}{\Sigma\omega_1\omega_2\omega_3\omega_4}e^{\!-\frac{2}{E}(\arccos\Sigma-\Sigma\sqrt{1-\Sigma^2})} \;,
\end{split}
\ee  
where the integrals are restricted to the region with $\omega_{1,2}>0$ and $\omega_{3,4}<0$, and $\Sigma=(\omega_1+\omega_2)/2$.
It is now straightforward to check that~\eqref{Gamma4onlydwleft} agrees with our WKB results for the amplitude. We again perform the momentum integral as before and find
\be
2\text{Im }\varepsilon^4\Gamma_4\approx V_3\int\frac{\ud^3p}{(2\pi)^3}\sum_\text{spin}|\varepsilon^2\frak{A}_2|^2 \;,
\ee
with $\frak{A}_2$ given by~\eqref{finalA2general}. (Note again that we have an approximate sign because we have neglected the subleading terms with $2\text{Re }\frak{A}_0^*\frak{A}_4$ and $2\text{Re }\frak{A}_1^*\frak{A}_3$.)

Thus, we have now obtained the same $P_3$ and $P_4$ using two completely different approaches, and without choosing a particular field shape of the weak field. For a Gaussian weak field~\eqref{WeakGaussianDef}, performing the remaining Fourier integrals with the saddle point method gives~\eqref{totalP3} and~\eqref{totalP4} (for $d=2$ transverse dimensions).

\begin{figure}
\includegraphics[width=0.95\linewidth]{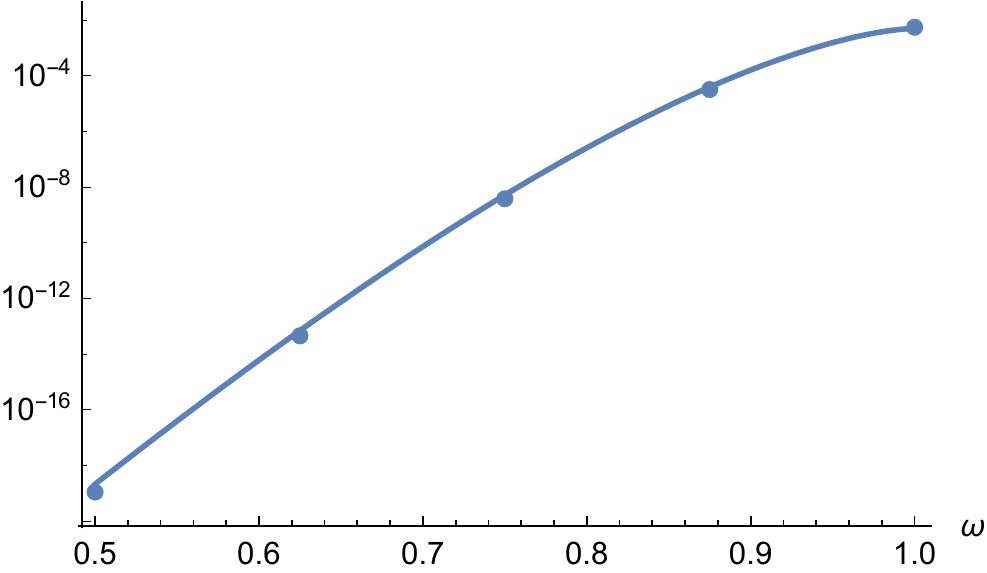}
\caption{$2\text{Im }\Gamma_4$ for $\omega_1=\omega_2=-\omega_3=-\omega_4=\omega$ and without the factor of $V_3\int\prod_{k=1}^4\left[\frac{\ud\omega_k}{2\pi}a(\omega_k)\right]2\pi\delta\left(\sum_{k=1}^4\omega_k\right)$. The dots are obtained by numerically integrating~\eqref{GammaExpandedEuclidean} and the line shows the analytical approximation~\eqref{Gamma4onlydwleft}, which is only valid for $\omega<1$ where the result is exponentially suppressed.}
\label{Gamma4numFig}
\end{figure}

The integrals in~\eqref{GammaExpandedEuclidean} can also be performed numerically. One approach is to first perform the $s$-integral by integrating along e.g. a C-shaped contour that passes vertically through the saddle point, which depends on $\tau_i$, or a similar contour in regimes where the result is not exponentially suppressed. Then one can perform the $\tau_i$ integrals on a real $N-1$ dimensional unit hypercube $0<\tau_i<1$. In Fig.~\ref{Gamma4numFig} we show the results of such a numerical integration for $\Gamma_4$ and $\omega_1=\omega_2=-\omega_3=-\omega_4=\omega$.
Of course, even for a monochromatic field we have $N$ integrals to perform for $\Gamma_N$, and the integrand becomes more complicated at higher $N$ because of the increase in the number of terms in the prefactor $W_N$, which can make a numerical integration time-consuming at high orders.

As a straightforward generalization of the above calculations we can also obtain higher orders. We already have the saddle points. What remains is to find some suitable integration variables and their scaling with respect to $E$, and then expand the integrand in $E$. 
We find exactly the same results as from the amplitude approach, i.e. $2\text{Im }\varepsilon^5\Gamma_5=\eqref{totalP5}$ and $2\text{Im }\varepsilon^6\Gamma_6=\eqref{totalP6}$. 

As yet another approach, we have also derived~\eqref{Gamma3onlydwlefttot} and~\eqref{Gamma4onlydwleft} by calculating the corresponding loop diagrams in Fig.~\ref{GammaExpansion1-fig} using the electron propagator in~\eqref{PropagatorInConstantE} (or rather the single-integral representation obtained by first performing the momentum integrals in~\eqref{PropagatorInConstantE}). The prefactor can then be obtained by choosing a representation for the Dirac matrices. This might at first seem like a simpler approach, but we found it much simpler to obtain~\eqref{Gamma3onlydwlefttot} and~\eqref{Gamma4onlydwleft} with the path-integral approach described in this section.

\section{Double assistance}
\label{Double assistance section}

So far we have considered a strong constant field assisted by a single weak field. In~\cite{Torgrimsson:2016ant} we proposed and studied a doubly assisted generalization, where the strong field is assisted by both a weak field~\cite{Schutzhold:2008pz} as well as a real/on-shell high-energy photon~\cite{Dunne:2009gi}. In~\cite{Torgrimsson:2016ant} we treated the weak field with nonperturbative methods. Here we will show that one can treat it with our perturbative approach. The inclusion of the high-energy photon basically corresponds to adding a third field in the shape of a plane wave, which is treated to lowest order.    
\begin{figure*}
\includegraphics[width=\textwidth]{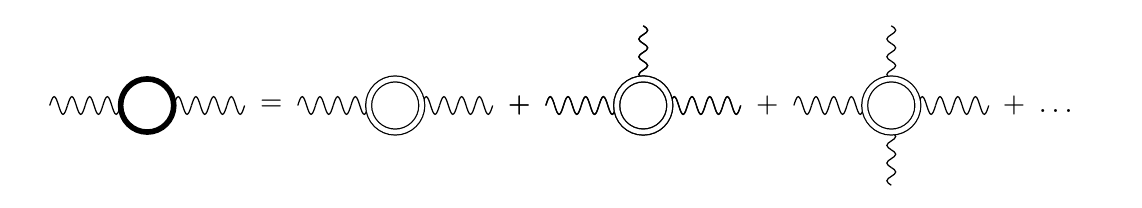}
\caption{The expansion of the polarization tensor. The bold and double lines again represent fermions dressed by both fields and only the strong field, respectively. The horizontal photon lines represent the single high-energy photon, and the vertical photon lines represent photons from the weak field. The pair production probability is obtained by applying the optical theorem.}
\label{polTensorExpanded-fig}
\end{figure*}
The pair production probability can be obtained from the polarization tensor using the optical theorem. Its weak field expansion is illustrated in Fig.~\ref{polTensorExpanded-fig}. The polarization tensor can be obtained from the following worldline representation of the effective action (see e.g.~\cite{Schubert:2001he,Schubert:2000yt,SchubertLectureNotes})
\be\label{spinor-Gamma}
\begin{split}
\Gamma&_{k,\epsilon\to k',\epsilon'}=2e^2\int_0^\infty\frac{\ud T}{T}\oint\mathcal{D}x\int\frac{\mathcal{D}\psi}{4}\int_0^1\ud\tau_1\ud\tau_2 \\
&\left[\epsilon\dot{x}+Tk\psi\epsilon\psi\right]_{\tau_1}\left[\epsilon'\dot{x}-Tk'\psi\epsilon'\psi\right]_{\tau_2}e^{-ikx(\tau_1)+ik'x(\tau_2)} \\
&\exp-i\left\{\frac{T}{2}+\int_0^1\frac{\dot{x}^2}{2T}+A\dot{x}-\frac{i}{2}\psi\dot{\psi}+\frac{i}{2}\psi TF\psi\right\} \;,
\end{split}
\ee 
where $\psi_\mu(\tau)$ is an anticommuting Grassmann variable with antisymmetric boundary conditions, $\psi(1)=-\psi(0)$, $F_{\mu\nu}=\partial_\mu A_\nu-\partial_\nu A_\mu$, and $k_\mu$ and $\epsilon_\mu$ are the momentum and polarization of the high-energy photon. We consider again $A_3=a(t)+Et$ and treat the weak field perturbatively using its Fourier transform~\eqref{aFourierDefinition}. This expansion makes the path integrals Gaussian and the prefactor is obtained from various Wick contractions as described in e.g.~\cite{Schubert:2001he,SchubertLectureNotes}; we have included the formulas we need in Appendix~\ref{Wick contractions in the worldline formalism}. The spatial homogeneity leads to the conservation of the photon momentum,
\be
\Gamma_{k,\epsilon\to k',\epsilon'}=:(2\pi)^3\delta^3(k'-k)iM_{\epsilon',\epsilon} \;.
\ee 
The optical theorem now gives the pair production probability
$P_{e^\LCp e^\LCm}=\frac{1}{k_0}\text{Im }M_{\epsilon,\epsilon}$.
For the high-energy photon we choose $k_\mu=\Omega(1,\sin\theta,0,\cos\theta)$ and two orthogonal polarization vectors $\epsilon^{(\LCpara)}_\mu=(0,-\cos\theta,0,\sin\theta)$ and $\epsilon^{(\LCperp)}_\mu=(0,0,1,0)$, which obey $k\epsilon=0$ and $\epsilon^2=-1$.

We focus on the perpendicular case, $k_3=0$, since this gives the largest probability and the simplest results. After performing the path integrals we find
\be
\begin{split}
\varepsilon^N P_N=&\text{Im}\int_0^\infty\!\ud T \int\prod_{i=1}^N\ud\omega_i a(\omega_i)\delta\left(\sum_{i=1}^N\omega_i\right)
\int\prod_{i=1}^{N+2}\ud\tau_i \\
&\dots e^{-i\left(\frac{T}{2}+\frac{1}{2}\sum_{k,l=1}^{N+2}K_k[\mathcal{G}_B(\tau_k-\tau_l)
-\mathcal{G}_B(0)]K_l\right)} \;,
\end{split}
\ee
where $K_{i,\mu}=\delta_\mu^0\omega_i$ for $i=1,..,N$, $K_{N+1,\mu}=k_\mu$, $K_{N+2,\mu}=-k'_\mu$, $\mathcal{G}_B$ is a worldline Green's function given by~\eqref{GBexplicit}, and where the ellipses stand for subdominant prefactor terms, see below, which are obtained from Wick contractions as described in~\eqref{pathAveGen}. 
We begin by finding the values of $\tau_i$ that maximize the exponential. This is similar to the case without the high-energy photon, and we again find that either $|\tau_i-\tau_j|=0$ or $|\tau_i-\tau_j|=1/2$. The $T$-integral is also similar to what we had in the previous sections. 
Using methods similar to the ones described above, we hence find
\be\label{PNexpDoublySigma}
\begin{split}
\varepsilon^NP_N\sim&\int\prod_{i=1}^N\ud\omega_i a(\omega_i)\delta\left(\sum_{i=1}^N\omega_i\right)\dots\\
&\exp\left\{-\frac{2m_\LCperp^2}{E}\left(\arccos\Sigma-\Sigma\sqrt{1-\Sigma^2}\right)\right\}  \;,
\end{split}
\ee
where $\Sigma$ is again the sum of the positive frequencies, but this time divided by an effective mass that depends on the frequency of the high-energy photon,
\be
\Sigma=\frac{1}{2m_\LCperp}\left(\Omega+\sum_{i=1}^J\omega_i\right)
\qquad
m_\LCperp^2=1+\left[\frac{\Omega}{2}\right]^2
\;,
\ee
where $0<J<N$ is an integer that characterizes different saddle points. For even $N$ the dominant contribution comes from $J=N/2$, and for a monochromatic field half of the Fourier frequencies must be positive implying $\sum_{i=1}^J\omega_i=N\omega/2$.
Compare~\eqref{PNexpDoublySigma} with~\eqref{PNintermsofSigma} for the case without the high-energy photon. The main difference is a heavy effective mass $m_\LCperp>1$ that comes from the spatial components of the high-energy photon momentum, which is similar to the results in~\cite{Torgrimsson:2017cyb} for singly assisted pair production with a weak field in the shape of a plane wave.   
Note that, even if the characteristic frequency $\omega_*$ of the weak field is much smaller than $\Omega$ and the electron mass, the dominant contributions for Gaussian and Sauter-like pulses still come from Fourier frequencies on the order of the electron mass $\omega_i\sim1$, similar to the case in the previous sections.

\subsection{Sauter pulse}

For a Sauter pulse~\eqref{SauterDef}, we find after performing the Fourier integrals
\be\label{finalExpSauter}
P_N\sim\exp\left\{-\frac{2m_\LCperp^2}{E}\left(-\frac{\Omega}{m_\LCperp\chi}+\frac{\sqrt{\chi^2-1}}{\chi^2}+\arcsin\frac{1}{\chi}\right)\right\} \;,
\ee
where $\chi=m_\LCperp\gamma_*$ and $\gamma_*=\omega_*/E$.
Note that all orders have the same exponential for these Sauter-like fields. That is what we found for ordinary dynamical assistance in~\cite{Torgrimsson:2017pzs}, and now we can see that this is also the case with the addition of a high-energy photon. 
Note also that~\eqref{finalExpSauter}, which is obtained by treating the weak field perturbatively, is exactly the same as the exponential we found in~\cite{Torgrimsson:2016ant} by treating the weak field nonperturbatively.

\subsection{Gaussian pulse}

For a Gaussian field~\eqref{WeakGaussianDef} the results are conveniently expressed in terms of $\nu=E/\omega^2$ and $\Lambda=\Omega/(2m_\LCperp)$. 
Performing the Fourier integrals with the saddle-point method leads to
\be\label{secondOrderGaussDoublyExp}
P_N\sim\exp\left\{-\frac{2m_\LCperp^2}{E}\left(\arccos\Sigma-\Lambda\bar{\nu}(\Sigma-\Lambda)\right)\right\}
\ee
where
\be
\Sigma=\frac{\bar{\nu}^2\Lambda+\sqrt{1+\bar{\nu}^2-\bar{\nu}^2\Lambda^2}}{1+\bar{\nu}^2}
\qquad
\bar{\nu}=\frac{N\nu}{2J(N-J)} \;.
\ee
The exponential is a strictly decreasing function of $\nu$ (which is natural since increasing $\nu$ corresponds to decreasing $\omega$). Thus, the dominant contribution comes from the value of $J$ that gives the smallest $\bar{\nu}$, which is $J=N/2$ for even $N$ and $J=(N\pm1)/2$ for odd $N$.
For $\Lambda\to0$ we recover our results for single assistance. For $\Lambda\ll1$ we have
\be
\Lambda\ll1: \qquad P_N\sim e^{-\frac{2m_\LCperp^2}{E}\left(\arctan\bar{\nu}-\frac{2\Lambda\bar{\nu}}{1+\bar{\nu}^2}\right)} \;,
\ee
which shows that the additional photon leads to a further reduction of the exponential suppression.
For $\nu\ll1$ the field strength drops out in the leading term in the exponent and we find for even $N$
\be
\nu\ll1:\qquad P_N\sim e^{-N\left(\frac{2m_\LCperp-\Omega}{N\omega}\right)^2\left(1-\frac{1}{3}[1-\Lambda]\bar{\nu}^2\right)} \;,
\ee
where the leading term is what one expects from $N$ factors of the Fourier transform evaluated at the minimum Fourier frequency needed to add up to the necessary energy, i.e. $(N/2)\omega_i=2m_\LCperp-\Omega$. 

As without the high-energy photon, the exponential increases while the prefactor decreases as we go to higher orders.
As in~\cite{Torgrimsson:2017pzs} we can estimate the probability by exponentiating $\varepsilon^N$ from the prefactor and approximating the sum over all orders with the ``saddle point'' for $N$, which we find to be
\be\label{NdomGauss}
N_{\rm dom}^{\rm Gauss}\sim2\nu\chi(\Sigma-\Lambda)
\quad\text{where}\quad
\Sigma=\sqrt{1-\frac{1}{\chi^2}} \;,
\ee $\chi:=\gamma_\LCperp/\sqrt{|\ln\varepsilon|}$ and $\gamma_\LCperp=m_\LCperp\gamma$. 
As $\Lambda\to0$ this reduces to the estimate in~\cite{Torgrimsson:2017pzs} of the dominant order in the singly assisted case. A nonzero $\Omega$ hence leads to a lower dominant order. Substituting the dominant order into $P_N$ gives us
\be\label{NdomExpGeneral}
P_{e^\LCp e^\LCm}^{\rm dom}\sim e^{-\frac{2m_\LCperp^2}{E}\left(-\frac{\Omega}{m_\LCperp\chi}+\frac{\sqrt{\chi^2-1}}{\chi^2}+\arcsin\frac{1}{\chi}\right)} \;.
\ee 
Curiously, this exponential has the same form as for a Sauter pulse~\eqref{finalExpSauter}, but with $\chi=\gamma_\LCperp/\gamma_{\rm crit}$ where $\gamma_{\rm crit}\sim\sqrt{|\ln\varepsilon|}$ in the Gaussian case. This generalizes a similar result in~\cite{Torgrimsson:2017pzs} to the case with an additional high-energy photon. A better agreement with the instanton exponent can be achieved by exponentiating a factor of $\gamma$ together with $\varepsilon$, so that $\gamma_{\rm crit}\to\sqrt{|\ln(c\varepsilon/\gamma)|}$, where $c$ is (to a first approximation) a constant obtained by matching, see~\cite{Torgrimsson:2017cyb}.
It might look like~\eqref{NdomExpGeneral} has a threshold at $\chi=1$, but $N_{\rm dom}^{\rm Gauss}>0$ (in~\eqref{NdomGauss}) implies $\chi>m_\LCperp$ so the threshold is given by $\gamma/\gamma_{\rm crit}=1$ and not $\gamma_\LCperp/\gamma_{\rm crit}=1$. We can also confirm this by noting that at $\chi=m_\LCperp$ the weak field drops out and we recover Eq.~(5) in~\cite{Dunne:2009gi}, which gives the exponential for the case where the strong constant field is only assisted by a high-energy photon.

\subsection{Sinusoidal field}\label{sinwt-section}

Our third example is a sinusoidal field $a(t)\propto\sin(\omega t)$. For this field we have
$\Sigma=\frac{1}{2m_\LCperp}\left(\Omega+\frac{N\omega}{2}\right)$. Estimating the dominant order as above we find results similar to the Gaussian case~\eqref{NdomGauss}, 
\be
N_{\rm dom}^{\rm cos}=\frac{4m_\LCperp}{\omega}(\Sigma-\Lambda) \quad\text{where}\quad
\Sigma=\sqrt{1-\frac{1}{\chi^2}} \;,
\ee
and
$\chi=\gamma_\LCperp/|\ln\varepsilon|$.
We again recover the result for the singly assisted case~\cite{Torgrimsson:2017pzs} as $\Omega\to0$. The threshold is again given by $\chi=m_\LCperp$. Substituting the dominant order into the exponential gives us~\eqref{NdomExpGeneral}, i.e. we again find the same form as in the Sauter case and the corresponding estimate for the Gaussian pulse, but with $\gamma_{\rm crit}^{\rm cos}\sim|\ln\varepsilon|$. 
We note that for $\gamma\gg\gamma_{\rm crit}$ we have
\be\label{sinwt-highgamma}
P_{e^\LCp e^\LCm}\sim\exp\left\{2\frac{2m_\LCperp-\Omega}{\omega}\ln\frac{\varepsilon}{\gamma}\right\} \;,
\ee  
which is simply the amplitude of the weak field $\varepsilon/\gamma$ to the power of the number of photons from the weak field that are needed to add up to twice the electron (effective) mass. 
   
To understand why we obtain~\eqref{NdomExpGeneral} for a sinusoidal field, notice that with $\hat{\omega}:=N\omega/2$ the sum over all orders $N$ can be expressed as
\be\label{NsumToomegaInt}
P_{e^\LCp e^\LCm}\sim\sum_{\hat{\omega}}e^{-2\frac{\hat{\omega}|\ln\varepsilon|}{\omega}-\frac{2m_\LCperp^2}{E}\left(\arccos\Sigma-\Sigma\sqrt{1-\Sigma^2}\right)} 
\ee 
where $\Sigma=(\Omega+\hat{\omega})/(2m_\LCperp)$,
so, by formally identifying $\hat{\omega}$ with the Fourier frequency in the second order case, we see that the $\ln\varepsilon$-term in~\eqref{NsumToomegaInt} behaves as the exponential decay~\eqref{SauterDef} of the Fourier transform of a Sauter pulse with an effective frequency $\omega_*=\omega/|\ln\varepsilon|$. Thus, estimating the sum in~\eqref{NsumToomegaInt} with the ``saddle point'' for $N$ leads to the Sauter exponential with $\gamma_{\rm crit}\sim|\ln\varepsilon|$.

\section{Conclusions}

This paper is a continuation of~\cite{Torgrimsson:2017pzs} where we study dynamically assisted Schwinger pair production by expanding the probability in a power series in the field strength of the weak field $\sim\varepsilon\ll1$. This approach allows us to obtain analytical approximations for a large class of fields, and hence provides a useful alternative to e.g. treating the total field with instanton methods. We can therefore learn more about the analytical structure of the probability, which is particularly important when assisting Schwinger pair production with high-energy photons. 

The Keldysh parameter of the weak field alone is large, $\omega/(\varepsilon E)\gg1$, and so the weak field is sometimes associated with the multiphoton regime. However, for weak fields with sufficiently wide Fourier transforms, like the exponentially decaying Fourier transform of a Sauter pulse, the dominant contribution comes already from the first order amplitude, $P_{e^\LCp e^\LCm}\sim|\varepsilon\frak{A}_1|^2$, i.e. from the absorption of a single photon. This means that both the exponential and the prefactor part of the probability can be calculated analytically for this class of fields~\cite{Torgrimsson:2017pzs}. For a Gaussian pulse the Fourier transform decays more rapidly and, although for some field parameters we still have $P_{e^\LCp e^\LCm}\sim|\varepsilon\frak{A}_1|^2$, in general one has to include higher orders in the $\varepsilon$ expansion, because the dominant contribution can come from one of the higher orders.
 
One of our main objectives in this paper is to show how to calculate the prefactor of higher-order terms in this expansion. We have showed how to use either WKB or worldline methods. We have for example derived the momentum spectrum using the worldline formalism~\cite{Dumlu:2011cc}. To the best of our knowledge, this is the first time that the preexponential factor of the momentum spectrum is derived using this formalism.  

As an example, we chose in~\cite{Torgrimsson:2017pzs} two sets of parameter values for a Gaussian field, one for which the exact/numerical results agree with $|\frak{A}_0+\frak{A}_1|^2$, and another with a weaker $E$ for which $|\frak{A}_0+\frak{A}_1|^2$ is clearly not enough. In this paper we have calculated $\frak{A}_2$ and showed that by including it we obtain a good approximation also for the second set of parameters. 
This agrees with our estimate of the dominant order~\cite{Torgrimsson:2017pzs}, see~\eqref{domNgauss}, which says that a weaker $E$ increases the dominant order.
This is an explicit example of the fact that, although $|\frak{A}_0+\frak{A}_1|^2$ is not enough for all field shapes or in all parameter regimes, one can nevertheless treat the weak field perturbatively, one just has to go to higher orders. Here we have obtained the prefactor up to $P_6$ (or $\frak{A}_3$), which was enough for a good approximation for the particular example just mentioned. In general the dominant contribution can of course come from even higher orders. It might become tedious at some point, but at least in principle one should be able to use the methods presented in this paper to obtain the prefactor of these higher orders as well.
In fact, as~\eqref{domNgauss} shows, the dominant order is mainly increased by a reduction of $E$, but a weaker $E$ also makes the probability much smaller (because of the exponential scaling), so for the most relevant parameter values one can quite generally expect the dominant order to still be low enough to not make the calculations impractical.

One advantage of our approach, where the weak field is expressed in terms of its Fourier transform, 
is that it becomes clear what frequency components are responsible for the dominant contribution.  
We have found that, e.g. for a Sauter pulse $\propto\text{sech}^2(\omega t)$ or Gaussian $\propto e^{-(\omega t)^2}$, the dominant contribution tends to come from Fourier frequencies on the order of the electron mass, even for $\omega\ll m$. 
If one insists on restricting the relevant frequencies to be below the electron mass, e.g. for experimental reasons, then one might be led to consider monochromatic fields, e.g. $\cos\omega t$. However, as the Fourier transform only has support at $\omega$, one then needs larger $\omega$, compared to the characteristic frequency of a Gaussian or a Sauter pulse, to obtain a significant enhancement, see e.g.~\cite{Torgrimsson:2017cyb}. So, in the parameter regime considered here it seems that for significant enhancement one is naturally led to consider frequencies that might be rather large compared to what near-future lasers can provide, but at least these higher frequencies make it easier to obtain simple approximations with the methods described here. 

In this paper we have focused on linearly polarized electric fields that only depend on time. This allows us to find simple, explicit analytical approximations. As shown in~\cite{Linder:2015vta,Schneider:2018huk}, purely time-dependent fields can, at least in some regimes, be used to give good quantitative approximations. It is also useful to start with such fields because it allows us to compare with the exact result obtained with well-developed numerical methods like solving the Riccati equation, which can be done to high precision~\cite{Schneider:2016vrl}, or the Wigner/quantum kinetic theory, which could be used for e.g. rotating fields~\cite{Blinne:2013via,Kohlfurst:2018kxg}.
However, our perturbative approach can also be useful for studying weak fields with more complex spacetime structure and/or strong fields with e.g. a nonzero magnetic component. For example, in~\cite{Torgrimsson:2017cyb} we applied our perturbative approach to a weak field in the shape of a plane wave, i.e. a case where the total field is an exact solution to Maxwell's equation in vacuum. We again found good agreement with results obtained with other methods. We found qualitatively similar behavior as for purely time-dependent fields, e.g. the existence of a dominant order, which provides further motivation for studying purely time-dependent electric fields.  
   
To further demonstrate the usefulness of this perturbative approach,
we have also applied it to doubly assisted pair production~\cite{Torgrimsson:2016ant}, where a high-energy photon is added to ordinary dynamical assistance. For Sauter-like weak fields we again find that the dominant contribution to the probability is quadratic in the weak field and its exponential part is exactly the same as the one obtained in~\cite{Torgrimsson:2016ant} by treating both the strong and the weak field with nonperturbative methods. As in the singly assisted case~\cite{Torgrimsson:2016ant}, we again find that a Gaussian or monochromatic weak field can lead to a higher dominant order. Although we have for simplicity assumed that both the (coherent) fields are purely time dependent, the high-energy photon is on-shell, so this is another multidimensional example, and here we have showed that it is still possible to calculate the prefactor.         

When extending the methods presented here to more complex, spacetime dependent fields, one might have to perform some steps numerically, e.g. to find the saddle points. Although the approximation would then not be completely analytical, one would still see the analytical dependence on some of the parameters and it could be very useful for quickly obtaining estimates in cases where an exact numerical treatment would be challenging or time-consuming. This could be useful for searching for promising parameter values for maximizing the enhancement of the probability for future experiments, before turning to a fully numerical treatment~\cite{Aleksandrov:2018uqb,Kohlfurst:2017git,Kohlfurst:2017hbd,Aleksandrov:2017mtq,Lv:2018wpn}. Moreover, as demonstrated in~\cite{Schneider:2018huk} and~\cite{Dinu:2018efz}, knowing the saddle points for some simpler fields can be very useful for finding the corresponding ones for complex fields that can be reached via a continuous deformation, which gives further motivation for working out all the details for simple fields as a start. 

\vspace{0.1cm}

\acknowledgements

G.~T. thanks Christian Schneider and Ralf Sch\"utzhold for many useful and inspiring discussions, and Christian Kohlf\"urst for useful discussions, especially about numerics, and for helpful comments on the draft.  
G.~T. is supported by the Alexander von Humboldt foundation.

\appendix

\section{Ingredients for the WKB approach}\label{Ingredients for the WKB approach}

In this appendix we collect some of the main ingredients needed in the WKB approach.
The WKB approximations are given by (see e.g.~\cite{Hebenstreit:2011pm,Hebenstreit:2010vz})
\be\label{UandV}
\begin{split}
U_r(t,{\bf q})&=(\gamma^0\pi_0+\gamma^i\pi_i+1)G^+(t,{\bf q})R_r \\
V_r(t,-{\bf q})&=(-\gamma^0\pi_0+\gamma^i\pi_i+1)G^-(t,{\bf q})R_r \;,
\end{split}
\ee
where $R_r$, $r=1,2$, are eigenspinors $\gamma^0\gamma^3 R_s=R_s$, and
\be\label{Gpmdef}
G^\pm(t,{\bf q})=[2\pi_0(\pi_0\pm\pi_3)]^{-\frac{1}{2}}\exp\bigg[\mp i\int_{t_0}^t\!\ud t'\,\pi_0(t')\bigg] \;,
\ee
where $\pi_3(t)=p_3-A(t)$ and $\pi_0=\sqrt{m_\LCperp^2+\pi_3^2(t)}$.
We arbitrarily choose $t_0=0$. These WKB approximations are eigenstates of the Hamiltonian (cf. e.g.~\cite{Fradkin:1991zq}) 
\be
\mathcal{H}=\gamma^0(-i\gamma^i\partial_i+\slashed{A}+1)
\ee  
\be
\mathcal{H}e^{-ip_ix^i}U(t,{\bf p})=\pi_0(t)e^{-ip_ix^i}U(t,{\bf p})
\ee
\be
\mathcal{H}e^{ip_ix^i}V(t,{\bf p})=-\pi_0(t)\Big|_{A\to-A}e^{ip_ix^i}V(t,{\bf p}) \;.
\ee
It follows from $\gamma^0\gamma^3 R_s=R_s$ that
$R_s^\dagger\gamma^0 R_r=R_s^\dagger\gamma^3 R_r=-(\gamma^3 R_s)^\dagger R_r=-R_s^\dagger\gamma^0 R_r=0$
and similarly $R_s^\dagger\gamma^0\gamma^\LCperp R_r=0$. Using these equations it is straightforward to show that 
\be
U_s^\dagger(t,{\bf q})U_r(t,{\bf q})=V_s^\dagger(t,{\bf q})V_r(t,{\bf q})=\delta_{sr}
\ee 
and 
\be
U_s^\dagger(t,{\bf q})V_r(t,-{\bf p})=0 \;.
\ee 

For a constant strong field $A=Et$, the integral in the exponent is given by
\be\label{intpi0tophi}
\int_0^t\pi_0=-\frac{m_\LCperp^2}{2E}\left(\phi\left[\frac{p_3-Et}{m_\LCperp}\right]-\phi\left[\frac{p_3}{m_\LCperp}\right]\right) \;,
\ee
where the second term is irrelevant and cancels upon squaring the amplitude to obtain the probability, and
\be
\phi(u)=u\sqrt{1+u^2}+\text{arcsinh }u \;.
\ee
For the first order amplitude we also readily find
\be
\bar{U}_s({\bf p})\gamma^3 V_{s'}(-{\bf p})=\delta_{ss'}\frac{m_\LCperp}{\pi_0}e^{\dots} \;.
\ee

\section{$\frak{A}_n$ from $\Gamma_n$}\label{AnFromGammanSec}

In this section we will show how to generalize the method in Sec.~\ref{Momentum spectrum from the worldline formalism} to higher orders. The idea is that to leading order we have
\be\label{AnFromGamman}
V_3\int\frac{\ud^3p}{(2\pi)^3}\sum_\text{spin}2\text{Re }\frak{A}_0^*\varepsilon^n\frak{A}_n=2\text{Im }\Gamma_n(\omega_i\omega_j>0) \;,
\ee
where $\Gamma_n(\omega_i\omega_j>0)$ is the contribution to $\Gamma_n$ in which all Fourier frequencies have the same sign. We can use~\eqref{AnFromGamman} to check that the methods in Sec.~\ref{Momentum spectrum from the worldline formalism} and~\ref{Using the propagator in a constant electric field} give the same results, but since $\frak{A}_0$ is so simple (see~\eqref{zerothOrderA}) we can actually use~\eqref{AnFromGamman} to extract $\frak{A}_n$ from $\Gamma_n$. This is useful because from $\frak{A}_n$ we obtain the dominant contribution to $P_{2n}$, and $\Gamma_n(\omega_i\omega_j>0)$ is simpler to calculate than $\Gamma_{2n}$. Note  that $\Gamma_n(\omega_i\omega_j>0)$ does not give the dominant contribution to $\Gamma_n$, which instead involves both positive and negative $\omega_i$. We calculate $\Gamma_n(\omega_i\omega_j>0)$ here in order to extract $\frak{A}_n$. 
The starting point is again~\eqref{worldline-spectrum-start}, which we expand in $\varepsilon$. This leads to three different factors in the pre-exponential part of the integrand,
\be
\int_0^1\!\ud\tau_j a'(t(\tau_j))=\int_0^1\!\ud\tau_j\int\frac{\ud\omega_j}{2\pi}a(\omega_j)(-i\omega_j)e^{-i\omega_jt(\tau_j)} \;,
\ee
\be\label{pmET}
\begin{split}
&\int_0^1\!\ud\tau_j[p_3-Et(\tau_j)]a(t(\tau_j))= \\
&\int_0^1\!\ud\tau_j\int\frac{\ud\omega_j}{2\pi}a(\omega_j)\left[p_3-iE\frac{\partial}{\partial\omega_j}\right]e^{-i\omega_jt(\tau_j)} 
\end{split}
\ee
and
\be\label{Esa2}
\begin{split}
\int_0^1\!\ud\tau_j\frac{E}{s}a^2(t(\tau_j))=
\int_0^1\!&\ud\tau_j\ud\tau_k\int\frac{\ud\omega_j}{2\pi}\frac{\ud\omega_k}{2\pi}a(\omega_j)a(\omega_k) \\
&\frac{E}{s}\delta(\tau_j-\tau_k)e^{-i\omega_jt(\tau_j)-i\omega_kt(\tau_k)} \;.
\end{split}
\ee
The $t$ path integral is now Gaussian and can be performed by removing the linear terms in the exponent with $t(\tau)\to t(\tau)+t_{\rm cl}(\tau)$ where $t_{\rm cl}$ is given by~\eqref{tclsum1}, and the resulting Gaussian integral gives~\eqref{tpathbeta}. The term in~\eqref{pmET} becomes
\be
\begin{split}
&\left[p_3-iE\frac{\partial}{\partial\omega_j}\right]e^{-\frac{i}{E}p_3\sum_{k=1}^n\omega_k} \\
&e^{-\frac{1}{4E}\sum_{k,l=1}^n\omega_k\omega_l\frac{\cos[s(1-2|\tau_k-\tau_l|)]}{\sin s}}=iEt_{\rm cl}(\tau_j)e^{...} \;.
\end{split}
\ee
The terms with e.g. $\partial t_{\rm cl}(\tau_j)/\partial\omega_k$ can be neglected to leading order in $E$. For $\omega_i\omega_j>0$ the exponent is maximized by $|\tau_i-\tau_j|=0,1$ for $i,j=1,...,n$. We substitute this into the prefactor and expand the exponent to leading order. This gives terms with $e^{-s\omega_i\omega_j|\tau_i-\tau_j|/E}$, which lead to elementary $\tau$ integrals. There is one $\tau$ integral that is trivial because of translation invariance. The other, nontrivial $\tau$ integrals each gives a factor of $E$, which means that the $E\delta(\tau_i-\tau_j)$ term in~\eqref{Esa2} is on the same order of magnitude as the other terms.
We now have the exponent in~\eqref{expPsSigmaCot}. We can therefore perform the $s$ integral in exactly the same way as for $n=1$. So, the saddle point is given by~\eqref{sSaddleGamma1} and the contribution to the prefactor is given by~\eqref{sSaddleIntegralGamma1}. The exponent is now given by
\be\label{A0Anexp}
e^{-\frac{m_\LCperp^2}{E}\frac{\pi}{2}}
e^{-\frac{m_\LCperp^2}{E}\left[2iP\Sigma+\arccos\Sigma-\Sigma\sqrt{1-\Sigma^2}\right]} \;.
\ee
The first part of the exponent comes from $\frak{A}_0$ (see~\eqref{zerothOrderA}) and the second part is the same as the one we obtained in Sec.~\ref{Using the propagator in a constant electric field} for $\frak{A}_n$ (see~\eqref{AnExpSigma}).

We now only have the Fourier integrals left, which we can perform with the saddle-point method for a Gaussian weak field. The exponential contribution from the Fourier transform depends on the $\omega_i$ variables separately, while the exponent in~\eqref{A0Anexp} only depends on their sum via $\Sigma=\sum\omega_i/(2m_\LCperp)$. One option is to free the $\Sigma$ variable so that we can use it as an integration variable, which can be achieved by inserting the following into the integrand 
\be
1=\int\ud\Sigma\int\frac{\ud\lambda}{2\pi}e^{i\lambda\left(\Sigma-\frac{1}{2m_\LCperp}\sum_{i=1}^n\omega_i\right)} \;.
\ee
It is now simple to perform the $\omega_i$ integrals with the saddle-point method, which gives a Gaussian $\lambda$ integral. Instead of introducing the $\lambda$ integral one can change variable e.g. from $\omega_1$ to $\Sigma$ and then perform the remaining $\omega_i$ integrals with the saddle-point method.  
The exponent is now given by
\be
e^{-\frac{m_\LCperp^2}{E}\frac{\pi}{2}}
e^{-\frac{m_\LCperp^2}{E}\left[\nu_n\Sigma^2+2iP\Sigma+\arccos\Sigma-\Sigma\sqrt{1-\Sigma^2}\right]} \;,
\ee
where $\nu_n=E/(n\omega^2)$.
We also perform the final integral with the saddle-point method. The saddle point for $\Sigma$ is given by~\eqref{SigmanDef}. The final exponent for the momentum spectrum is given by
\be\label{eeSigman}
e^{-\frac{m_\LCperp^2}{E}\frac{\pi}{2}}
e^{-\frac{m_\LCperp^2}{E}\left[iP\Sigma_n+\arccos\Sigma_n\right]} \;,
\ee
where $\Sigma_n$ is given by~\eqref{SigmanDef}. It is now straightforward to obtain the prefactor. We just multiply together the contributions from the $\tau$ integrals and the Gaussian integrals around the saddle points for the $s$ and $\omega_i$ integrals, and substitute $|\tau_i-\tau_j|=0$, $s=\frac{\pi}{2}+\text{arccos}\Sigma_n$, $\omega_i=2m\Sigma_n/n$ and $\Sigma_n$ from ~\eqref{SigmanDef} into the rest of the prefactor. For $n=2$ we find~\eqref{Gamma2w1w2SameSign} and~\eqref{A2Gauss}. 
For $n=3$ we find
\be
\begin{split}
2\text{Im }\Gamma_3(\omega_i\omega_j>0)=&V_3\int\frac{\ud^3p}{(2\pi)^3}e^{-\frac{\pi m_\LCperp^2}{2E}}4\text{Re} \\
&\left[\frac{E\varepsilon}{\omega}\right]^3\frac{27\sqrt{3\pi}E}{128m_\LCperp^5\Sigma_3^8\nu_3}\frac{9-8\Sigma_3^2}{\sqrt{1+\nu_3^2+\frac{i\nu_3P}{\Sigma_3}}} \\
&e^{-\frac{m_\LCperp^2}{E}\left[iP\Sigma_3+\arccos\Sigma_3\right]} \;.
\end{split}
\ee
From this we can immediately extract the third order amplitude $\frak{A}_3$ using~\eqref{AnFromGamman} and~\eqref{zerothOrderA}, and the result is the same as the one we obtained in~\eqref{A3Gauss} with the propagator approach.

\section{Higher orders for a Gaussian pulse}\label{higherOrderAnGaussian}

After we have performed the Fourier integrals, the exponent in the amplitude is given by~\eqref{eeSigman}
\be
\frak{A}_n\sim e^{-\frac{m_\LCperp^2}{E}\left[iP\Sigma_n+\arccos\Sigma_n\right]} \;.
\ee
Now we can integrate $\frak{A}_m^*\frak{A}_n$ over the momentum with the saddle-point method. The saddle point for the longitudinal momentum, $P_{nm}$, is determined by $\Sigma_n(P_{nm})=\Sigma_m(-P_{nm})$, which leads to a purely imaginary (or zero for $m=n$) solution given by
\be\label{Psaddlenm}
P_{nm}=i\frac{\nu_n-\nu_m}{\sqrt{4+(\nu_n+\nu_m)^2}} \;.
\ee
Substituting~\eqref{Psaddlenm} into~\eqref{SigmanDef} gives
\be
\Sigma_n(P_{nm})=\left[1+\left(\frac{\nu_n+\nu_m}{2}\right)^2\right]^{-\frac{1}{2}} \;.
\ee    
The perpendicular momentum integrals are dominated by $p_\LCperp=0$.
Substituting these saddle points into the exponent we finally obtain
\be\label{finAmAnexp}
\int\!\ud^3p\;\frak{A}_m^*\frak{A}_n\sim\exp\left\{-\frac{2}{E}\arctan\frac{\nu_n+\nu_m}{2}\right\} \;.
\ee
Consider the $N$-th order of the probability $P_N$. The amplitudes that contribute to this have $m=N-n$ and hence 
\be\label{finalPN}
P_N\sim\sum_{n=0}^N\dots\exp\left\{-\frac{2}{E}\arctan\frac{N\nu}{2n(N-n)}\right\} \;.
\ee
This is exactly the same as the exponents we found in~\cite{Torgrimsson:2017pzs} using a very different approach, see Eq.~(5.10) and~(5.11) in~\cite{Torgrimsson:2017pzs}. In~\cite{Torgrimsson:2017pzs} we obtained this exponential from the worldline representation of the effective action or the master formulas for $N$-photon scattering in~\cite{Schubert:2000yt}. Those approaches give directly the total/integrated probability with no reference to the amplitude or any momentum integrals. By rederiving this exponential with the current approach, we learn that the different saddle points we found in~\cite{Torgrimsson:2017pzs}, which are characterized by $n$ in~\eqref{finalPN}, correspond to the products of the different amplitude orders, $\frak{A}_{N-n}^*\frak{A}_n$, that contribute to the probability $P_N$ at a given order. For even $N$ we see that the largest contribution comes from $n=N/2$, and for odd $N$ the largest contribution comes from $n=(N\pm1)/2$, i.e. (cf.~Eq.~(3.7) in~\cite{Torgrimsson:2017pzs})
\be
\begin{split}
	\text{N even:} \qquad P_N&\sim|\frak{A}_{N/2}|^2\sim\exp\left\{-\frac{2}{E}\arctan\frac{2\nu}{N}\right\} \\
	\text{N odd:} \qquad P_N&\sim 2\text{Re }\frak{A}_{(N-1)/2}^*\frak{A}_{(N+1)/2} \\
	&\sim\exp\left\{-\frac{2}{E}\arctan\frac{2N\nu}{N^2-1}\right\} \;.
\end{split}
\ee
As we go to higher orders, $\varepsilon^N$ in the prefactor decreases while the exponential increases, which leads in general to the existence of a dominant order~\cite{Torgrimsson:2017pzs}.

\begin{widetext}

\section{Wick contractions in the worldline formalism}
\label{Wick contractions in the worldline formalism}

To obtain the prefactor for the doubly assisted case, we have used different methods. In one of them the spin factor is expressed in terms of a Grassmann path integral and the prefactor is obtained from Wick contractions. There are well-known techniques, see e.g.~\cite{Schubert:2001he}, for calculating such Wick contractions in arbitrary constant fields. We collect here the results we need in our case. The basic ingredients are the worldline Green's functions, $\mathcal{G}_B$ and $\mathcal{G}_F$, for the $x$ and $\psi$ path integrals, respectively. 
Let $g^{\scriptscriptstyle\parallel}_{\mu\nu}
=\delta_\mu^0\delta_\nu^0-\delta_\mu^3\delta_\nu^3$, $g_{\mu\nu}^\LCperp=-\delta_\mu^1\delta_\nu^1-\delta_\mu^2\delta_\nu^2$ and 
%$F_{\mu\nu}=E\hat{F}_{\mu\nu}$, where 
$\hat{F}_{\mu\nu}=\delta_\mu^0\delta_\nu^3-
\delta_\mu^3\delta_\nu^0$. 
The bosonic Green's function is given by 
\be\label{GBexplicit}
\begin{split}
\mathcal{G}^B_{\mu\nu}(\tau,\tau')
=&g^\LCperp_{\mu\nu}T\left(\frac{1}{2}[|\tau-\tau'|-(\tau-\tau')^2]-\frac{1}{12}\right)+ \\
&g^{\scriptscriptstyle\parallel}_{\mu\nu}\frac{-i}{2E}\left(\frac{\cos[s(1-2|\tau-\tau'|)]}{\sin s}-\frac{1}{s}\right)+\hat{F}_{\mu\nu}\frac{\epsilon(\tau-\tau')}{2E}\left(\frac{\sin[s(1-2|\tau-\tau'|)]}{\sin s}-(1-2|\tau-\tau'|)\right) \;,
\end{split}
\ee
where $s=iET/2$.
We have $\mathcal{G}^B_{\mu\nu}(\tau,\tau')=\mathcal{G}^B_{\nu\mu}(\tau',\tau)$, $\mathcal{G}^B_{\mu\nu}(1,\tau')=\mathcal{G}^B_{\mu\nu}(0,\tau')$ and
$\left(\frac{\partial_\tau^2}{T}-F\partial_\tau\right)\mathcal{G}_B(\tau,\tau')=\delta(\tau-\tau')-1$ (the identity matrix is the Minkowski one, $1_{\mu\nu}\to g_{\mu\nu}$).
The fermionic Green's function is given by
\be\label{GFexplicit}
\begin{split}
\mathcal{G}^F_{\mu\nu}(\tau-\tau')
=g^\LCperp_{\mu\nu}\frac{\epsilon(\tau-\tau')}{2}+g^{\scriptscriptstyle\parallel}_{\mu\nu}\frac{\epsilon(\tau-\tau')}{2}\frac{\cos[s(1-2|\tau-\tau'|)]}{\cos s}+\hat{F}_{\mu\nu}\frac{i}{2}\frac{\sin[s(1-2|\tau-\tau'|)]}{\cos s} \;,
\end{split}
\ee
which satisfies $\mathcal{G}^F_{\mu\nu}(\tau,\tau')=-\mathcal{G}^F_{\nu\mu}(\tau',\tau)$, $\mathcal{G}^F(1,\tau')=-\mathcal{G}^F(0,\tau')$ and
$(\partial_\tau-TF)\mathcal{G}^F(\tau,\tau')=\delta(\tau-\tau')$. 
These Green's functions are the Minkowski versions of the Euclidean ones in e.g.~\cite{Schubert:2000yt,Schubert:2001he}. 

We have integrals in the form
\be\label{generalxpathintegrals}
\int\mathcal{D}x\prod_{i=1}^I\eta_{b_i}\dot{x}(\tau_{b_i})\exp\left\{-i\int_0^1\frac{\dot{x}^2}{2T}+Et\dot{z}+jx\right\} \;,
\ee
where $1\leq b_i,I\leq N$, $\eta^\mu$ is the polarization vector of either the high-energy photon ($\epsilon,\epsilon'$) or the weak field ($a(\omega_i)$), and
\be
j_\mu=k_\mu\delta(\tau-\tau_{N+1})-k'_\mu\delta(\tau-\tau_{N+2})+\delta_\mu^0\sum_{k=1}^N\omega_k\delta(\tau-\tau_k)=:\sum_{k=1}^{N+2}K_{k,\mu}\delta(\tau-\tau_k) \;.
\ee
We begin by integrating over the center of mass, $x^\mu(\tau)\to x_{\rm cm}^\mu+x^\mu(\tau)$ where $\int_0^1 x=0$, which gives delta functions. Next we exponentiate each $\eta\dot{x}$ factor and then perform the resulting Gaussian integrals as described in Sec.~\ref{Momentum spectrum from the worldline formalism} and~\ref{First order Gamma1}.  
We thus find
\be
\eqref{generalxpathintegrals}=(2\pi)^3\delta^3(k-k')2\pi\delta\left(\sum_{k=1}^N\omega_k\right)\text{lin}_\eta\exp\left\{-\frac{i}{2}\int J\mathcal{G}_BJ\right\}\frac{1}{(2\pi iT)^2}\frac{s}{\sin s} \;,
\ee
where (cf.~\cite{Schubert:2000yt})
\be\label{JGBJ}
\int J\mathcal{G}_BJ=\sum_{k,l=1}^NK_k[\mathcal{G}_B(\tau_k-\tau_l)-\mathcal{G}_B(0)]K_l-2iK_k\dot{\mathcal{G}}_B(\tau_k-\tau_l)\eta_l+\eta_k\ddot{\mathcal{G}}_B(\tau_k-\tau_l)\eta_l \;,
\ee
and $\text{lin}_\eta$ selects the terms that are linear in all the $\eta_{b_i}$ that appear in the prefactor of~\eqref{generalxpathintegrals} (the other $\eta$'s in this sum are zero).

For the Grassmann path integral we find
\be 
\int\frac{\mathcal{D}\psi}{4}\prod_{r=1}^Rv_r\psi(\tau_{f_r})\exp\left\{-\int_0^1\frac{1}{2}(\psi_0\dot{\psi}_0-\psi_i\dot{\psi}_i)-ET\psi_3\psi_0\right\}
=\text{lin}_\xi\exp\left\{\frac{1}{2}\sum_{r,r'=1}^R\xi_r^\mu\xi_{r'}^\nu\mathcal{G}^F_{\mu\nu}(\tau_{f_r}-\tau_{f_{r'}})\right\}\cos s \;,
\ee
where $v_{r,\mu}$ is either $k$, $\kappa$, $\epsilon$, $a(\omega_i)$ etc, $f_r$ is an integer, $1\leq f_r\leq N$, and where $\xi_{r,\mu}=v_{r,\mu}\xi_r$ are Grassmann valued and $\text{lin}_\xi$ selects the terms that are proportional to $\xi_1\xi_2...\xi_R$ (the order is important since they are anticommuting). The contractions come in pairs with two equal $\tau$'s (e.g. $\tau_{f_1}=\tau_{f_2}=\tau_1$). 
 
Thus, the Wick contractions we need can be obtained from
\be\label{pathAveGen}
\begin{split}
&\left\langle\prod_{i=1}^I\eta_{b_i}^\mu\dot{x}_\mu(\tau_{b_i})
\prod_{r=1}^Rv_r^\mu\psi_\mu(\tau_{f_r})\right\rangle \\
&=\text{lin}_{\eta,\xi}\exp\left\{\sum_{k,l=1}^N\left(-K_k^\mu\dot{\mathcal{G}}^B_{\mu\nu}(\tau_k-\tau_l)\eta_l^\nu-\frac{i}{2}\eta_k^\mu\ddot{\mathcal{G}}^B_{\mu\nu}(\tau_k-\tau_l)\eta_l^\nu\right)+\frac{1}{2}\sum_{r,r'=1}^R\xi_r\xi_{r'}v_r^\mu\mathcal{G}^F_{\mu\nu}(\tau_{l_r}-
\tau_{l_{r'}})v_{r'}^\nu\right\} \;,
\end{split}
\ee
where $\eta_i^\mu$ and $v_i^\mu$ etc are the same as above.

\subsection{Prefactor for double assistance}

Here we will consider the prefactor for double assistance to second order in the weak field. Our starting point is
\be\label{Gamma2start}
\begin{split}
M_{\epsilon,\epsilon'}^{(2)}=2e^2\int_0^\infty&\frac{\ud T}{T}\frac{s\cot s}{(2\pi i T)^2}\int\frac{\ud\omega_1}{2\pi}\frac{\ud\omega_2}{2\pi}2\pi\delta(\omega_1+\omega_2)\int\limits_0^1\ud\tau_1
\ud\tau_2\ud\tau_3\ud\tau_4\Big\langle \frac{-1}{2}\left[a\dot{x}+T\kappa\psi a\psi\right]_{\omega_1,\tau_1}\left[a\dot{x}-T\kappa\psi a\psi\right]_{\omega_2,\tau_2} \\
&\left[\epsilon\dot{x}+Tk\psi\epsilon\psi\right]_{\tau_3}\left[\epsilon'\dot{x}
-Tk'\psi\epsilon'\psi\right]_{\tau_4}\Big\rangle
\exp-i\left(\frac{T}{2}+\frac{1}{2}\sum_{k,l=1}^NK_k[\mathcal{G}_B(\tau_k-\tau_l)
-\mathcal{G}_B(0)]K_l\right) \;,
\end{split}
\ee 
where $K_1=\kappa$, $K_2=-\kappa$, $K_3=k$, $K_4=-k'$, and $\kappa_\mu=\omega_1\delta_\mu^0$. The factor of $-1/2$ comes from expanding the exponential in~\eqref{spinor-Gamma} to second order in the weak field. 
The Wick contractions in $\langle...\rangle$ are obtained from~\eqref{pathAveGen}, and the integrals are performed with the saddle-point method or generalizations thereof, as explained above. We find for high-energy photons with parallel and perpendicular polarization 
\be\label{PparaPperpDoubly}
\begin{split}
P_{\scriptscriptstyle\parallel,\perp}=\frac{\alpha E}{\Omega}\int\frac{\ud\omega_1}{2\pi}\frac{|a(\omega_1)|^2}{\omega_1^2\Sigma}\left\{\frac{4\sqrt{1-\Sigma^2}}{\Omega^2},\frac{8(1-\Sigma^2)+\omega_1^2}{4m_\LCperp^2\sqrt{1-\Sigma^2}}\right\}&\left[\arccos\Sigma\left(\arccos\Sigma-\frac{p_1^2}{m_\LCperp^2}\frac{\sqrt{1-\Sigma^2}}{\Sigma}\right)\right]^{-\frac{1}{2}} \\
&\exp\left\{-\frac{2m_\LCperp^2}{E}\left(\arccos\Sigma-\Sigma\sqrt{1-\Sigma^2}\right)\right\} \;,
\end{split}
\ee
where $\Sigma=(\Omega+\omega_1)/(2m_\LCperp)$, $m_\LCperp=\sqrt{1+p_1^2}$ and $p_1=\Omega/2$. 
This prefactor can also be obtained using Feynman's path-ordered representation of the spin factor. 
A third option is to use the WKB approach, i.e. by basically just replacing one $\slashed{a}$ in~\eqref{A2start} with $\slashed{\epsilon}e^{-ikx}$, and then following the same steps as before. It turns out that for this process the WKB approach actually allows us to obtain the prefactor with less effort than the worldline approach, because it is easier to calculate the prefactor using an explicit Dirac matrix representation than to calculate Grassmann Wick contractions.

\end{widetext}

\end{document}